\documentclass[12pt,preprint]{aastex}
\usepackage{apjfonts}

\newcommand\qc[1]{}
\newcommand\ac[1]{}

\def\Rs{R_{\rm S}}

\newcommand\bmath[1] {\mbox{\boldmath$\rm #1$}}

\newcommand\s{{\rm s}} 
\newcommand\hr{{\rm hr}} 
\newcommand\yr{{\rm yr}} 
\newcommand\GHz{{\rm GHz}} 
\newcommand\m{{\rm m}} 
\newcommand\mm{{\rm m}\m} 
\newcommand\Ms{M_\odot} 
\newcommand\erg{{\rm erg}} 
\newcommand\G{{\rm G}} 
\newcommand\e{{\rm e}}

\begin{document}
%
\slugcomment{Version 4.01: \today}
%
\shorttitle{Sgr A* Flare Detection with Millimeter VLBI}
\shortauthors{Doeleman et al.}
\title{Methods for Detecting Flaring Structures in Sagittarius A* with High Frequency VLBI}
\author{
Sheperd S.~Doeleman\altaffilmark{1},
Vincent L.~Fish\altaffilmark{1},
Avery E.~Broderick\altaffilmark{2},
Abraham Loeb\altaffilmark{3}, \&
Alan E.~E.~Rogers\altaffilmark{1}}
\altaffiltext{1}{Massachusetts Institute of Technology, Haystack
  Observatory, Route 40, Westford, MA 01886.}
\altaffiltext{2}{Canadian Institute for Theoretical Astrophysics,
  University of Toronto, 60 St.\ George St., Toronto, ON, M5S 3H8
  Canada.}
\altaffiltext{3}{Institute for Theory and Computation, Harvard
  University, Center for Astrophysics, 60 Garden St., Cambridge, MA
  02138.}

\begin{abstract}
The super massive black hole candidate, Sagittarius~A*, exhibits variability
from radio to X-ray wavelengths on time scales that correspond to < 10
Schwarzschild radii.  We survey the potential of millimeter-wavelength VLBI to
detect and constrain time variable structures that could give rise to such
variations, focusing on a model in which an orbiting hot spot is embedded in an
accretion disk.  Non-imaging algorithms are developed that use interferometric
closure quantities to test for periodicity, and applied to an ensemble of
hot-spot models that sample a range of parameter space.  We find that
structural periodicity in a wide range of cases can be detected on most
potential VLBI arrays using modern VLBI instrumentation.  Future enhancements
of mm/sub-mm VLBI arrays including phased array processors to aggregate VLBI
station collecting area, increased bandwidth recording, and addition of new
VLBI sites all significantly aid periodicity detection.  The methods described herein
can be applied to other models of Sagittarius~A*, including jet outflows and
Magneto-Hydrodynamic accretion simulations.
\end{abstract}
\keywords{black hole physics --- Galaxy: center --- techniques:
  interferometric --- submillimeter --- accretion, accretion disks}

\section{Introduction}
\label{introduction}

Observations of the compact radio/IR/X-ray source Sagittarius~A*
(Sgr~A*) make the most compelling case for the existence of super massive
black holes.  Both speckle imaging and adaptive optics work in the
near-infrared (NIR) band shows that multiple stars orbit the position
of Sgr~A* \citep{schodel03,ghez05} These orbits are consistent with a
central mass of $\sim4\times10^6 ~ M_\sun$ contained within 45~AU -
the closest approach of any star.  Radio interferometric proper motion
measurements \citep{backer99,reid04} limit the motion of Sgr~A* to
$<15$~km\,s$^{-1}$, implying that Sgr~A* must trace at least 10\% of
the mass determined from stellar orbits.  Very long baseline interferometry (VLBI) at 1.3~mm
wavelength \citep{doeleman08} has resolved Sgr~A*, and measures an intrinsic
size of $<0.3$~AU \citep[assuming a distance of 8.0~kpc, from][]{reid93}, or 4
times the Schwarzschild radius of the central mass ($\Rs \approx 10~\mu$as).
Assuming Sgr~A* marks the position of the black hole, the implied density,
using the proper motion lower limit on the mass and the VLBI size, is
$>9.3\times10^5~M_\sun\,{\mathrm{AU}}^{-3}$.  Almost any conceivable
aggregation of matter would, at this density, quickly collapse to a black hole
\citep{maoz98}.

The 1.3~mm VLBI result confirms the existence of structures within Sgr~A* on
size scales commensurate with the innermost accretion region, and matches size
scales inferred from light curve monitoring over a broad wavelength range.
Sgr~A* exhibits variability on time scales of minutes to hours in the radio,
millimeter, NIR, and X-ray bands \citep[e.g.,][]{baganoff01, aschenbach04,
genzel03, ghez04, belanger06, meyer06, yusefzadeh06, hornstein07, marrone07},
and flare rise times in the X-ray and NIR correspond to light-crossing times of
$<12~\Rs$.  Models that produce flaring X-ray flux density via Synchrotron Self
Compton scattering of IR photons require emission regions of diameter $<10~\Rs$
\citep{marrone07}.  

The resolution of millimeter- and submillimeter-wavelength VLBI is
well matched to the scale of inner disk physics.  Baselines from
Hawaii or Western Europe to Chile provide fringe spacings as small as
$30~\mu$as ($3~\Rs$) at 230~GHz and $20~\mu$as at 345~GHz.  Millimeter VLBI
thus has the potential to detect signatures of hot spot and jet models proposed
to explain the rapid variability of Sgr~A* as well as strong general
relativistic effects, such as the black hole silhouette or shadow
\citep{falcke00,broderick06b,huang07,markoff07}.  Extending the VLBI technique
to short (0.85~mm, 1.3~mm) wavelengths is essential for this work due to the
interstellar scattering towards Sgr~A*, which broadens radio images with a
$\lambda^2$ dependence \citep{backer78}.  VLBI at 7~mm and 3~mm wavelengths
\citep{rogers94,doeleman01,bower04,shen05} limits the intrinsic size of Sgr~A*
to be $<2$~AU and $<1$~AU respectively, but VLBI at these wavelengths is
strongly influenced by scattering effects.  For $\lambda < 1.3$~mm the scattering size is
less than the fringe spacings on the longest possible baselines.  Recent 1.3~mm
VLBI results \citep{doeleman08}, coupled with ongoing technical advances to
reach 0.85~mm, strongly suggest that it is not a question of \emph{if} but of
\emph{when} VLBI will directly probe Sgr~A* on event horizon scales.

Claims of observed periodicity in IR and X-ray light curves
\citep{belanger06, meyer06, eckart06} during Sgr~A* flares, can
potentially be explained in the context of hot spots orbiting the
black hole at a few times $\Rs$.  It has been proposed that the
fastest periodicity can be used to constrain the spin of the black
hole, since the period of the innermost stable circular orbit (ISCO)
is much shorter for a maximally rotating Kerr black hole than for a
nonrotating Schwarzschild black hole \citep{genzel03}.  Indeed, several
authors have argued that the black hole must be rotating based on
observed rapid X-ray and infrared periodicities
\citep{aschenbach04,belanger06,meyer06}.  However, it has also been
proposed that the flaring activity can be explained by
magnetohydrodynamic turbulence along with density fluctuations
\citep{goldston05,chan06}.  Rossby wave instabilities may naturally
produce periodicities on the order of tens of minutes, in which case
it is not necessary to appeal to a nonzero black hole spin to explain
the fast quasi-periodic flares seen at multiple wavelengths
\citep{tagger06,falanga07}.  Regardless of the source of variability
at millimeter wavelengths, VLBI has the potential to confirm
conclusively its association with the inner disk of Sgr~A*, to probe
the size of the region of variability (since an interferometer acts as
a spatial filter on the emission), and to extract periodicity.

Preliminary studies involving the analysis of simulated data from
expected models are important for a number of reasons.  Such studies
will highlight the abilities and limitations of millimeter VLBI in
regards to detecting the signatures of the physical processes in the
accretion disk surrounding the black hole in Sgr~A*.  Critical
resources (such as stable frequency standards, high-bandwidth
recording equipment, and phased array processors) are likely to be
limited initially, and telescope upgrades (such as surface accuracy
improvement, expanded IF bandwidth, additional receiver bands, and
simultaneous dual-polarization capability) must necessarily be prioritized.
Simulated observational data can help assess the tradeoffs that must be
considered for optimization of Galactic Center VLBI observations.  The ultimate
goal is to explore the potential of black hole parameter estimation by present
and future millimeter VLBI observations.

In this manuscript, we explore the observational signatures of an orbiting hot
spot embedded in a quiescent disk around Sgr~A*.  We consider a non-imaging
approach to analyzing millimeter VLBI data for several reasons.  First, one
fundamental assumption of Earth rotation aperture synthesis is that the source
structure is not changing.  Since orbital periods in the hot spot models are
much shorter than the rotation of the Earth, this assumption is clearly
violated.  Second, phase-referencing the data is presently not feasible at
millimeter wavelengths since the phase path through the atmosphere changes on a
time scale which is faster than the time needed to move the antennas between
the reference source and Sgr A*.  We note, though, that this problem could be
circumvented at connected-element arrays where some antennas could be dedicated
to simultaneously observing a reference source while others observe Sgr A*, but
this is also currently limited by the low SNR that can be achieved in the
coherence time of the atmosphere.  Potential arrays for millimeter VLBI will
have a small number of telescopes, initially as few as three, possibly with
vastly different sensitivies.  Low expected signal-to-noise ratios (SNRs) on
some baselines combined with the few antennas available will prevent adequate
self-calibration via closure relations.  Third, even if the visibility data
could be properly calibrated and the source structure were not changing over
the observation, the $(u,v)$-plane would be sparsely populated with noisy data
points, resulting in poor image fidelity, as shown by the simulations of
\citet{miyoshi04}.  At least initially, it will be more productive to analyze
the data by model fitting in the visibility domain rather than in the image
domain.

\section{Models of Sgr A*}
\label{models}

Models for the flaring emission of Sgr~A* at millimeter wavelengths
necessarily require a number of components, most succinctly decomposed
into models for the quiescent emission and models for the
short-timescale dynamical phenomena responsible for the flare.  Any
such model has a number of existing observational constraints that it
must meet, including reproducing the observed spectra \& polarization
properties of the quiescent \& flaring emission as well the dynamical
properties of the flare light curves.  Here we describe a set of flare
models involving orbiting hot spots embedded within a large-scale
accretion flow that are consistent with all existing observations,
based upon those described in \citet{broderick06b}.

\subsection{Quiescent Emission}

As implied by its spectrum, Sgr~A* is only starting to become optically thin at
millimeter wavelengths.  Due to relativistic effects this does not
happen isotropically \citep[e.g.,][]{broderick06a}.  As a consequence,
the opacity of the underlying accretion flow is important for both
imaging the black hole's silhouette and for the variability arising
from hot spots on compact orbits.

Despite being faint compared to the Eddington luminosity for
a $4\times10^6\,\Ms$ black hole, Sgr~A* is still considerably bright,
emitting a bolometric luminosity of approximately $10^{36}\,\erg\,
\s^{-1}$.  As a consequence it has been widely accepted that Sgr~A*
must be accretion powered, implying a minimum accretion rate of at
least $10^{-10}\,\Ms\, \yr^{-1}$.

It is presently unclear how this emission is produced.  This is
evidenced by the variety of models that have been proposed to explain
the emission characteristics of Sgr A*
\citep[e.g.,][]{narayan98,blandford99,falcke00,yuan02,yuan03,loeb07}.
Models in which the emission arises directly from the accreting
material have been subsumed into the general class of Radiatively
Inefficient Accretion Flows (RIAF), defined by the generally weak
coupling between the electrons, which radiate rapidly, and the ions,
which efficiently convert gravitational potential energy into heat
\citep{narayan98}. This coupling may be sufficiently weak to allow
accretion flows substantially in excess of the $10^{-10}\,\Ms\,
\yr^{-1}$ required to explain the observed luminosity with a canonical
radiative efficiency.

Nevertheless, following the detection of polarization from Sgr A*
above $100\,\GHz$ \citep{aitken00,bower01,bower03,marrone06}, and
subsequent measurements of the Faraday rotation measure
\citep{macquart06,marrone07b}, the accretion rate near the black hole
has been inferred to be significantly less than the Bondi rate,
implying the existence large-scale outflows \citep{agol00,quataert00}.
Therefore, in the absence of an unambiguous theory, we adopt a simple,
self-similar model for the underlying accretion flow which includes
substantial mass-loss.

Following \citet{yuan03}, this model is
characterized by a Keplerian velocity distribution, a population of
thermal electrons with density and temperature
\begin{equation}
n_{e,\rm th} = n^0_{e,\rm th} \left(\frac{r}{\Rs}\right)^{-1.1} \e^{-z^2/2 \rho^2}
\quad {and} \quad
T_{e} = T^0_{e} \left(\frac{r}{\Rs}\right)^{-0.84}\,,
\end{equation}
respectively, a population of non-thermal electrons
\begin{equation}
n_{e,\rm nth} = n^0_{e,\rm nth} \left(\frac{r}{\Rs}\right)^{-2.9} \e^{-z^2/2 \rho^2}\,,
\end{equation}
and spectral index $\alpha_{\rm disk} = 1.25$ (defined as
$S\propto\nu^{-\alpha_{\rm disk}}$), and a toroidal magnetic field in
approximate ($\beta=10$) equipartition with the ions (which produce
the majority of the pressure), i.e.,
\begin{equation}
\frac{B^2}{8\pi}
=
\beta^{-1} n_{e,\rm th} \frac{m_p c^2 \Rs}{12 r}\,.
\end{equation}
In all of these expressions the radial structure was taken directly
from \citet{yuan03} and the vertical structure was determined by
assuming the disk height is comparable to the polar radius, $\rho$.
To correct for the fact that \citet{yuan03} was a Newtonian study, we
determine the three coefficients ($n^0_{e,\rm th}$, $T^0_e$ and
$n^0_{e,\rm nth}$) by fitting the the radio, submillimeter and
near-infrared spectrum of Sgr~A*.  For every inclination and black
hole spin presented here this was possible with extraordinary accuracy
(reduced $\chi^2<1$ in all cases and $\lesssim0.2$ for many), implying
that this model is presently significantly under-constrained by the
quiescent spectrum alone.  It is also capable of producing the Faraday
rotation measures observed, and thus the polarimetric properties of
Sgr~A*.

The primary emission mechanism is synchrotron, arising from both the
thermal and non-thermal electrons.  We model the emission from the
thermal electrons using the emissivity described in \citet{yuan03},
appropriately altered to account for relativistic effects \citep[see,
  e.g.,][]{broderick04}.  Since we perform polarized radiative
transfer via the entire complement of Stokes parameters, we employ the
polarization fraction for thermal synchrotron as derived in
\citet{petrosian83}.  In doing so we have implicitly assumed that the
emission due to thermal electrons is isotropic, which while generally
not the case is unlikely to change our results significantly.  For the
non-thermal electrons we follow \citet{jones77} for a power-law
electron distribution, cutting the electron distribution off below a
Lorentz factor of $10^2$ and corresponding to a spectral index of
$\alpha_{\rm disk}=1.25$, both roughly in agreement with the
assumptions in \citet{yuan03}.  For both the thermal and non-thermal
electrons the absorption coefficients are determined directly via
Kirchoff's law.

\subsection{Flares}

We model the flare emission by a localized over-density in the
non-thermal electron distribution.  This naturally explains the
short-timescale variability and potential periodicity claimed in some
flares \citep{genzel03,belanger06}.  Such a feature is also a natural
consequence of dissipation in black hole accretion flows.

Sgr~A*'s quiescent radio spectrum appears to require a population of
non-thermal electrons.  If strong magnetic turbulence is present,
driven by, e.g., the magnetorotational instability, the production of
non-thermal electrons at strong shocks and magnetic reconnection
events is unavoidable.  Generally, we expect that the production of
non-thermal electrons will be most prominent in the innermost regions
of the accretion flow, where the magnetic turbulence is strongest.

It is important to note that the region producing the flare need not
be dynamically important.  Within the context of a RIAF model for the
accretion flow onto Sgr~A*, the pressure is overwhelmingly dominated
by the ions.  This is a direct consequence of the assumed weak
coupling between the electrons and ions, and thus the low luminosity
of the accretion flow.  For typical RIAF accretion rates
($10^{-8}\,\Ms\, \yr^{-1}$), the luminosity would need to be increased
by orders of magnitude before the non-thermal electrons become
dynamically significant.  For the observed flares, which typically do
not increase the NIR luminosities by more than a single order of
magnitude, this means that the accelerated electrons will be frozen
into the accretion flow.  In this case, the size of the emitting
region, $2 \Delta r$, is determined by the scale of the magnetic
turbulence.

The dominant constraints upon the lifetimes of hot spots in the
accretion flow are Keplerian shear and synchrotron cooling.  Hot spots
will shear apart on roughly $r/\Delta r$ orbital periods, and thus
small spots will last many orbits.  Unlike shear, cooling is not
achromatic, with the flare cooling more rapidly at higher frequencies.
The cooling time is approximately
\begin{equation}
\tau_{\rm c}
\simeq
3 \left(\frac{\lambda}{1\,\mm}\right)^{1/2} \left(\frac{B}{30\,
  \G}\right)^{-3/2}\,\hr\,,
\label{eqn-cooling}
\end{equation}
which should be compared to the period at the ISCO, ranging from
$30\,\min$ for a non-rotating black hole to $4\,\min$ for a maximally
rotating black hole.  Thus we expect that hot spots will typically
survive many orbits at millimeter and submillimeter wavelengths.

Our hot-spot model consists of a locally spherically symmetric,
Gaussian over-density of non-thermal electron.  Explicitly, given
$\Delta x^\mu = x^\mu-x_{\rm spot}^\mu$ and a scale, $R_{\rm spot}=0.75\Rs$,
the hot-spot density is
\begin{equation}
n_{e,\rm spot} = n^0_{e,\rm spot}
\exp\left[-\frac{\Delta x^\mu \Delta x_\mu + \left(u_{\rm spot}^\mu
    \Delta x_\mu\right)^2}{2 R_{\rm spot}^2} \right],
\end{equation}
where the hot-spot four-velocity, $u_{\rm spot}^\mu$ is assumed to be
the same as that of the underlying disk (which we have chosen to be
Keplerian).  Our description of the hot spot is completed by the
spectral index of the power-law distribution of electrons,
$\alpha_{\rm spot}=1.3$, taken from observations of NIR
flares \citep{eckart04}.  Like the disk, the hot-spot radiates
primarily via synchrotron.

\subsection{Generating Images}

The method by which images of the flaring disk are produced is
discussed at length in \citet{broderick03} \citep[see
  also,][]{broderick06a,broderick06b}.  As a result, we only briefly
summarize the procedure here.

Null geodesics are constructed by integrating a Hamiltonian
formulation of the geodesic equations:
\begin{equation}
\frac{dx^\mu}{d\eta} = f(x^\sigma) \, k^\mu
\quad{\rm and}\quad
\frac{dk_\mu}{d\eta} = - \left.\frac{f(x^\sigma)}{2}\frac{\partial k^\nu
  k_\nu}{\partial x^\mu}\right|_{k_\alpha}\,,
\end{equation}
where $f(x^\sigma)$ is an arbitrary function, corresponding to the
freedom inherent in the affine parametrization, $\eta$.  In order to
regularize the affine parametrization near the horizon, we choose
\begin{equation}
f(x^\sigma) = r^2\sqrt{1-\frac{r}{r_{\rm h}}},
\quad{\rm where}\quad
r_{\rm h } = \frac{\Rs}{2}\left( 1 + \sqrt{1-a^2} \right)
\end{equation}
is the horizon radius ($a$ is the dimensionless black hole
spin).  It can be explicitly shown that this does
reproduce the null geodesics \citep{broderick03}. 

The relativistic generalization of the radiative transfer problem is
most easily obtained by directly integrating the Boltzmann equation
directly \citep{lindquist66,broderick06a}.  In this case, it is the
photon distribution function, $N_\nu\propto I_\nu/\nu^3$ (which has
the virtue of being a Lorentz scalar), that is evolved.  In the case
of polarized transfer, it is possible to define the covariant
analogues of the Stokes parameters, ${\bf N}_\nu = \left( N_\nu,
N^Q_\nu, N^U_\nu, N^V_\nu \right)$, in terms of a parallel propagated
tetrad \citep{broderick04}.  In terms of these, the radiative transfer
equation takes on its standard form:
\begin{equation}
\frac{d{\bf N}_\nu}{d\eta}
=
{\bf \bar{j}}_\nu - \bmath{ \bar{\alpha}}_\nu {\bf N}_\nu\,,
\end{equation}
where ${\bf \bar{j}}$ and $\bmath{\bar{\alpha}_\nu}$ are the
appropriately generalized emission and absorption coefficients, and
may be trivially related to the same quantities in the local plasma
frame \citep{lindquist66,broderick04}.

Images are produced by tracing a collection of initially parallel null
geodesics from pixels on a distant plane backwards in time, towards
our model of Sgr~A*, terminating the ray when it has been captured by
the black hole, escaped the system, or accrued an optical depth
greater than $10$.  Along each ray we integrate the polarized
radiative transfer, obtaining ${\bf N}_\nu$ at the original plane.  We
construct 100 such images with resolutions of $128\times128$ pixels
for each hot-spot orbit.  This procedure is repeated for each
frequency of interest, for which we keep the underlying physical model
fixed.  Thus, for each spin/inclination pair, the relationship between
230~GHz and 345~GHz images is dictated by the spectral properties of
the source.

\section{VLBI Analysis}
\label{VLBI}

\subsection{Techniques}
\label{techniques}

Typical VLBI analysis techniques employ an iterative
``self-calibration'' loop, whereby a sky brightness model, the VLBI
data, and a series of complex gains for each VLBI site are brought
into convergence \citep{cornwell81,cornwell82}.  This process relies
on the assumption that all array calibration can be expressed as
station-based gains, which is equivalent to requiring that all
calibration errors 'close', or cancel when computed over suitable
closed loops of VLBI baselines.  In almost all cases, this assumption
is valid, and closure quantities can be constructed from the data,
which contain structural information on the observed source, but that
are largely immune to station-based phase and gain errors
\citep{jennison58,twiss60}.  However, closure quantities alone are
insufficient to determine baseline phases for a VLBI array.  The
number of independent closure phases computed over closed triangles of
VLBI stations, for example, grows as $\onehalf \,(N-1)\,(N-2),$ where
$N$ is the number of antennas, while the number baseline phases in the
array will be $\onehalf \, N \,(N-1).$ The fraction of visibility
phase information available from closure phases is thus $(N-2)/N$.
For millimeter VLBI $N$ will be small (initially $N = 3$ or $4$; \S
\ref{antennas}), and closure quantities will not be sufficiently
numerous to allow for full calibration of the data.  However, closure
quantities are robust observables and can therefore be used for model
fitting even when baseline-based visibilities are contaminated with
station-based phase and gain errors.  Indeed, closure quantities have
been successfully used for model fitting \citep[e.g.,][]{rogers74},
including recent experiments to place limits on the apparent size and
structure of Sgr~A* at wavelengths as short as 3~mm
\citep{doeleman01,bower04,shen05,markoff07}.

Closure phase and amplitude in the weak-detection limit are discussed
in detail by \citet{rogers95}.  We summarize the most relevant
information below.

The closure phase is the sum of the baseline phases along a triangle
of antennas.  It is independent of instrumental and atmospheric
complex gain terms.  The closure phase of a symmetric distribution of
emission is always zero or 180\degr\ \citep[e.g.,][]{monnier07}.
Deviations from these values are indicative of asymmetries (about the
origin) in the source structure on the size scales probed by the
baselines of the triangle.  A static asymmetric source structure will
show slow variations in closure phase over the course of observations
due to the rotation of the Earth, which changes the projected length
and orientation of baselines.  If the source structure is changing, as
would be the case for an orbiting hot spot, the closure phase on some
triangles will change as well.  Because the time scale of hot spot
orbits near the black hole is on the order of tens of minutes
(depending on the mass and spin of the black hole and the orbital
radius of the hot spot), closure phase variability will be much more
rapid for the case of a hot spot embedded in a disk as compared to a
quiescent disk alone.  Since a hot spot may survive for
several orbits before cooling or shearing (\S \ref{models}), closure
phases can exhibit approximate perodic behavior over several cycles.
Thus, closure phases are appropriate observables for detecting
periodic source structure changes on short time scales.  The SNR of
the closure phase is dominated by the lowest SNR of the visibility
data along the three segments.

The closure amplitude is constructed as the ratio of products of
visibility amplitudes ($A$) along a quadrangle of antennas.  For four
antennas $a$, $b$, $c$, and $d$, two independent closure amplitudes
can be constructed:
\begin{equation}
A_{abcd} \equiv \frac{|A_{ab}| |A_{cd}|}{|A_{ac}| |A_{bd}|}
\quad \mathrm{and} \quad
A_{adbc} \equiv \frac{|A_{ad}| |A_{bc}|}{|A_{ac}| |A_{bd}|}.
\end{equation}
For an array of $N$ antennas, there are $\onehalf\,N\,(N-3)$ independent
closure amplitudes.  As for closure phase, the closure amplitude is
unaffected by gain calibration errors on each of the antennas.
Deviations of the closure amplitude from unity are usually indicative
of resolved, asymmmetric source structure.  Large excursions in
closure phase and amplitude are often correlated, since small changes
with time in the complex visibility, due to Earth rotation or source
structure changes, have the largest effects on the phase when the
visibility amplitude is near zero.  Since a small visibility amplitude
results in a very small closure amplitude if it appears in the
numerator or a very large closure amplitdue if it appears in the denominator.
For this reason, closure amplitudes are generally presented on logarithmic
scales.

In the low-signal regime, the closure amplitude is a biased quantity.
The detected visibility amplitude on a baseline is the modulus of the
vector (amplitude and phase) sum of the signal and the noise and is by
definition nonnegative.  When the SNR is small, the visibility
amplitude is dominated by the noise amplitude and is therefore larger
than the signal amplitude.  Thus, if the source is not detected on
baseline $ab$ but is detected on baselines $cd$, $ac$, and $bd$, the
closure amplitude $A_{abcd}$ will on average be larger than the value
predicted by a noiseless model of the source structure.  The closure
phase does not suffer from a similar bias.

\subsection{Antennas}
\label{antennas}

\begin{deluxetable}{lrclrlr}
\tablewidth{\hsize}
\tablecaption{Assumed Telescope Parameters\label{tab-ants}}
\tablehead{
  \colhead{} &
  \colhead{} &
  \colhead{Diameter\tablenotemark{a}} &
  \colhead{230} &
  \colhead{SEFD\tablenotemark{b}} &
  \colhead{345} &
  \colhead{SEFD\tablenotemark{b}} \\
  \colhead{Facility} &
  \colhead{$N$\tablenotemark{a}} &
  \colhead{(m)} &
  \colhead{GHz?} &
  \colhead{(Jy)} &
  \colhead{GHz?} &
  \colhead{(Jy)} 
}
\startdata
 Hawaii &  8 & 23 &  yes &  4900   & yes  &  8100   \\
 CARMA  &  8 & 27 &  yes &  3500   &planned&  4900  \\
 SMT   &  1 & 10 &  yes & 11900   & yes  & 23100   \\
 LMT    &  1 & 32&planned& 10000\tablenotemark{c}& no& 13700\tablenotemark{d} \\
 ASTE   &  1 & 10 &  no  & \nodata & yes  & 14000   \\
 APEX   &  1 & 12 & soon &  6500   & yes  & 12200   \\
 ALMA   & 10 & 38 & soon &   500   & soon &  1000   \\
 PV     &  1 & 30 &  yes &  2900   & no   &  5200\tablenotemark{d} \\
 PdB    &  6 & 37 &  yes &  1600   & soon &  3400   
\enddata
\tablenotetext{a}{Effective aperture when number of antennas ($N$) are
  phased together.}
\tablenotetext{b}{Expected system equivalent flux density values at
  230 and 345~GHz toward Sgr~A* include typical weather conditions and
  opacities.}
\tablenotetext{c}{Completion of the dish and upgrades to the surface
  accuracy and receiver will eventually lower the SEFD by more than a
  factor of 10.}
\tablenotetext{d}{These telescopes do not presently have planned
  345~GHz capability.  Assumed SEFD values are for illustrative
  purposes only.}
\end{deluxetable}

\begin{figure}                                                                                                          
\resizebox{\hsize}{!}{\includegraphics{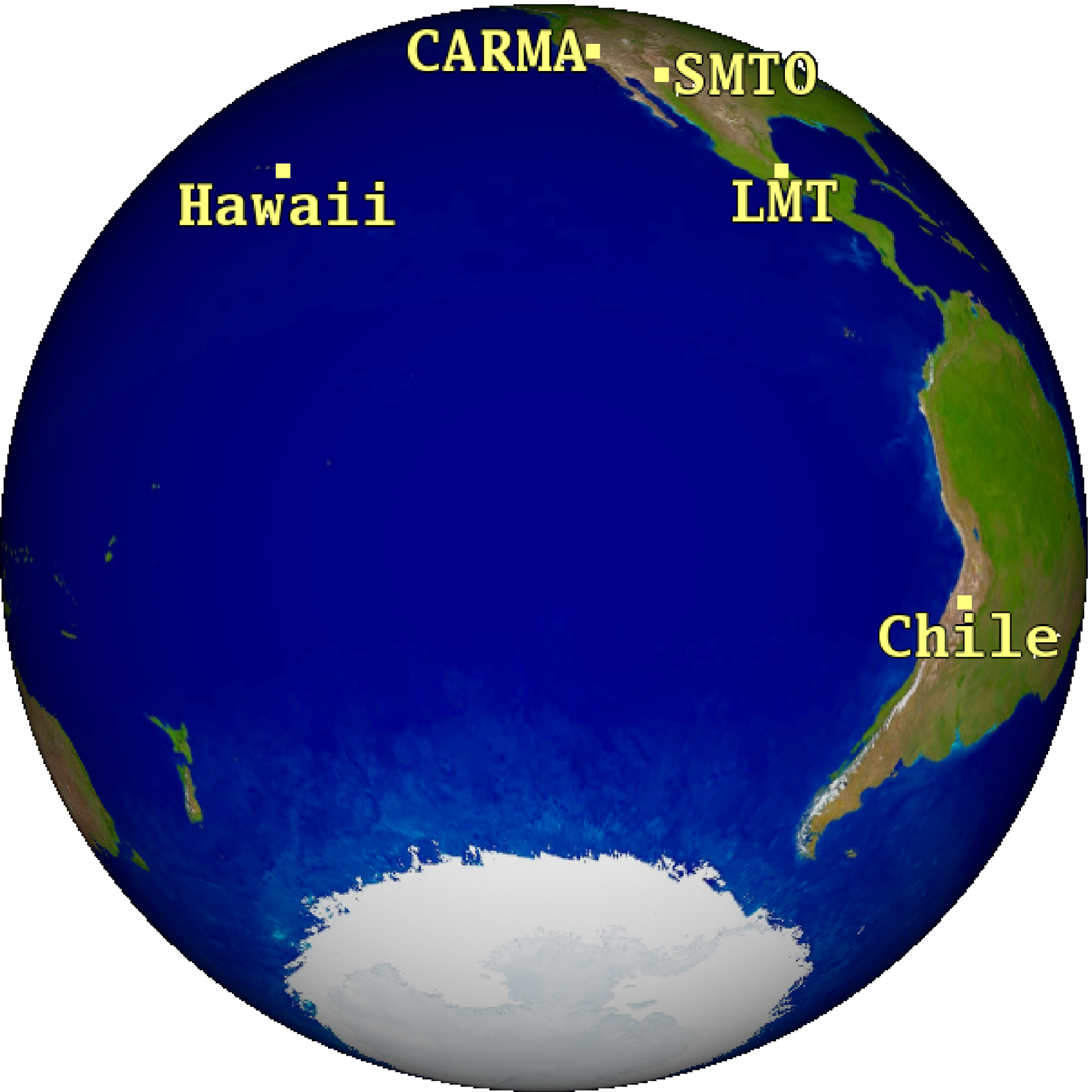}\includegraphics{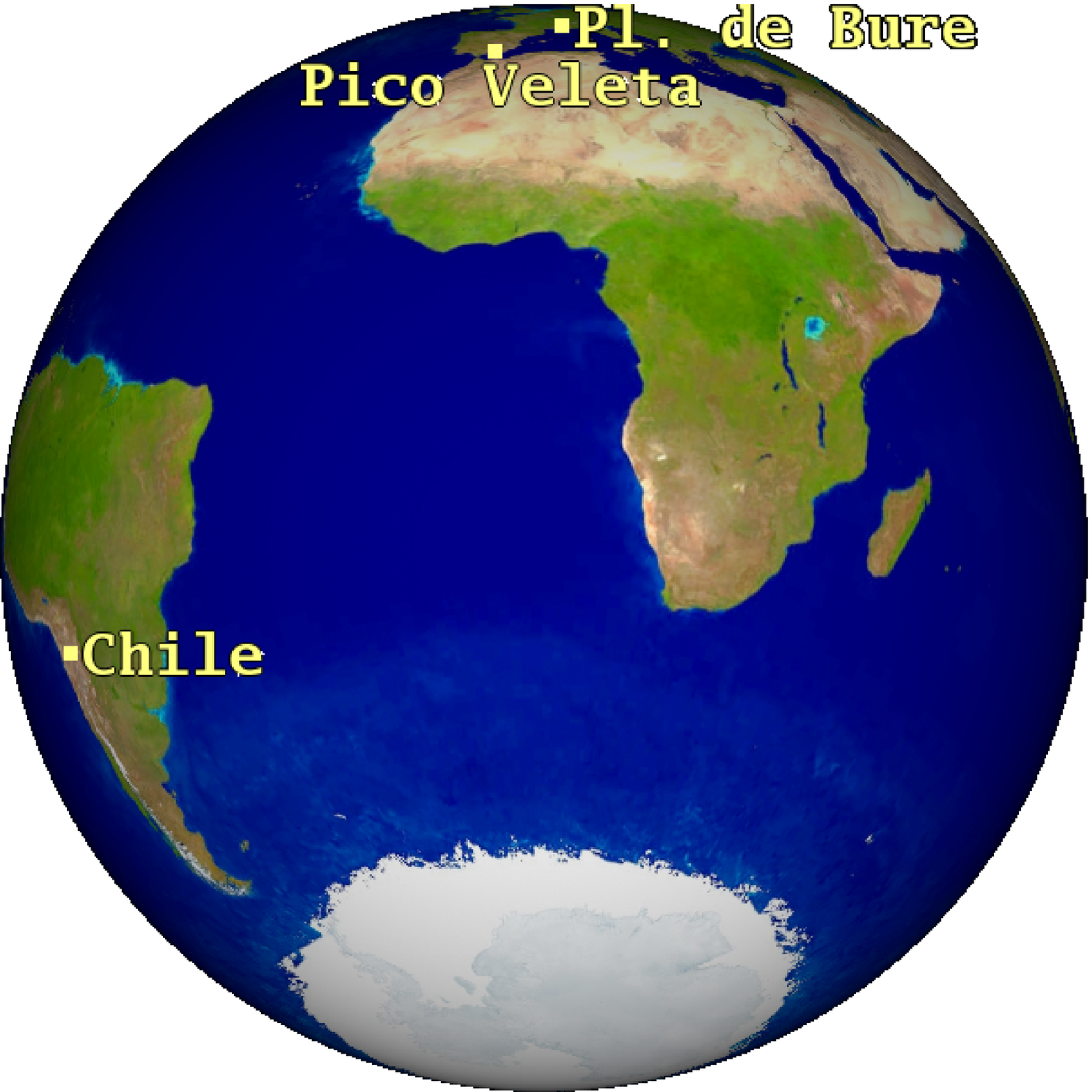}}                                               
\caption{Locations of candidate telescopes for millimeter-wavelength                                                    
  VLBI as viewed from the declination of Sgr~A*.                                                                        
\label{fig-locations}                                                                                                   
\notetoeditor{For the reader's convenience, Figure 1 should be placed                                                   
  near Table 1, preferably on the same page}                                                                            
}                                                                                                                       
\end{figure}             

Whereas centimeter-wavelength astronomy can be performed from
virtually any location with a clean spectrum, potential sites for
(sub)millimeter-wavelength astronomy are limited by the need to be
above most of the water vapor content in the atmosphere.
Consequently, possible arrays for millimeter VLBI are sparser
than for centimeter VLBI.  In this section, we summarize the best
prospects for millimeter VLBI among existing telescopes and those that
may come on line in the near future.

\emph{Hawaii}: The Caltech Submillimeter Observatory (CSO), James
Clerk Maxwell Telescope (JCMT), and Submillimeter Array (SMA) are all located atop Mauna
Kea.  Each telescope can observe in both the 230 and 345~GHz bands.
The JCMT has a single-polarization 230~GHz receiver and a
dual-polarization 345~GHz receiver.  The SMA presently does not
support simultaneous dual-polarization observations but is expected to
do so eventually at 345~GHz.  Development of instrumentation to phase
together the CSO, JCMT and several SMA dishes is underway, and will
result in an effective 23~m aperture, which we refer to as ``Hawaii,''
by early 2009.

\emph{CARMA}: The Combined Array for Research in Millimeter-wave
Astronomy (MWA) consists of six 10.4~m antennas and nine 6.1~m
antennas.  The ability to phase together up to 8 CARMA dishes
for VLBI is planned, and will result in an effective 27~m aperture.
CARMA can presently observe in the 230~GHz band.  A future upgrade to
include 345~GHz is planned, but it is unclear when this band will be
available for observations.  We also consider a single 10.4~m dish at
230~GHz (``CARMA 1'') in \S \ref{results-arrays}.

\emph{SMTO}: The Arizona Radio Observatory Submillimeter Telescope (SMT) on Mt.
Graham in Arizona is a 10~m dish capable of observing at both 230 and 345~GHz.

\emph{LMT}: The Large Millimeter Telescope (LMT), presently under
construction, will be a 50~m dish capable of observing at both the 230
and 345~GHz bands.  When complete, it will be the most sensitive
millimeter telescope in the Northern hemisphere.  In anticipation that 
the LMT collecting area will be installed in phases, we conservatively adopt
an effective aperture of 32~m.

\emph{Chile}: Several millimeter telescopes are available in Chile.
The Atacama Submillimeter Telescope Experiment (ASTE) is a 10~m dish
with a single-polarization receiver at 345~GHz.  The Atacama
Pathfinder Experiment (APEX) is a 12~m dish with a double-sideband
(DSB) at 345~GHz.  Future two sideband (2SB) heterodyne receivers
capable of observing at 230~GHz and 345~GHz are under construction.
We refer to using one of these facilities as ``Chile 1.''  ALMA will
be composed of a large number of 12~m dishes with 2SB receivers at
both 230 and 345~GHz.  We also consider the possibility of a
10-element phased ALMA station as the Chilean site (``Chile 10''),
since our models do not produce much expected signal on the long
baselines to Chile (\S \ref{methods}).

\emph{Pico Veleta}: The 30~m Institut de Radioastronomie
Millim\'{e}trique (IRAM) telescope on Pico Veleta (PV) can observe at
230~GHz.

\emph{Plateau de Bure}: The IRAM PdB Interferometer consists of six
15~m telescopes equipped with 230~GHz receivers.  An upgrade to add
345~GHz capability is presently under construction and is expected to
be available in late 2008.  The six telescopes can currently be phased
up over a 256~MHz bandwidth into a single 37~m-equivalent aperture.
Extension to higher bandwidths will require instrumentation
development similar to what will be deployed to phase up the antennas
on Mauna Kea.

A summary of telescope capabilities is given in Table~\ref{tab-ants}.
Sensitivity is given in terms of the system equivalent flux density
(SEFD), which is equal to $2 k T_\mathrm{sys}/A_\mathrm{eff}$, where
$k$ is Boltzmann's constant and the effective area $A_\mathrm{eff}$ is
the product of the geometric area and the aperture efficiency.
Assumed $T_\mathrm{sys}$ values toward Sgr~A* include a typical
expected atmospheric contribution, which can be quite large for
Northern hemisphere telescopes due to the low elevation of the
Galactic Center.  Since the atmospheric contribution is highly
weather-dependent, actual observations may achieve significantly
different values of the SEFD at each telescope.  SEFD values include a
10\% phasing loss factor for phased arrays.  All facilities can in
principle provide an IF bandwidth of at least 4~GHz, although
accessing the full bandwidth at some stations may be problematic.
Thus, 32~Gbit\,s$^{-1}$, corresponding to 2-bit Nyquist samples of a
4~GHz bandwidth in two orthogonal polarizations, is the maximum
possible recording rate that we will consider.  At present
16~Gbit\,s$^{-1}$ digital back ends are still in the planning stage,
so initial observations will likely employ a smaller bandwidth.
Except as noted, all telescopes can observe two polarizations
simultaneously.

The European telescopes (PV and PdB) and the North American telescopes
(Hawaii, CARMA, SMTO, and LMT) have no mutual visibility of Sgr~A*,
except for approximately 1~hr of overlap above 10\degr\ elevation on
the PV-LMT baseline.  Thus, there are only three possible types of
subarrays with at least three elements possible among these millimeter
facilities: North America only, North America plus Chile, and Europe
plus Chile (Fig.~\ref{fig-locations}).  For our simulations, we take our 230~GHz array to consist
of Hawaii, CARMA, SMTO, LMT (32~m), either APEX or a 10-element phased
ALMA, PV, and PdB.  Early science at 345~GHz will likely consist of
the single triangle of Hawaii, SMTO, and a Chilean telescope.
However, we consider the same set of telescopes as at 230~GHz in order
to illustrate what future 345~GHz upgrades might accomplish.

\subsection{Methods}
\label{methods}

Simulated data were obtained using task UVCON in the Astronomical
Image Processing System (AIPS).  Synthetic data were produced using an
averaging interval of 10~s.  Typical coherence times at 230~GHz are 10~s, but
can be as low as 2-4~s and, under good weather conditions, as long as 20~s
\citep{doeleman02}.  At the ALMA site in Chile, the measured coherence time of
the atmosphere is > 10 seconds 60\% of the time at 230GHz and 45\% of the time
at 345GHz \citep{holdaway97}.  Since AIPS cannot directly handle time-varying
source structure, for a hot spot model we simulate data from 100 static,
equally-spaced time slices in a single hot spot orbit and construct a data set.
All data assume 2-bit sampling at the Nyquist rate.  Thus, the recording rate
(in Gbit\,s$^{-1}$) is four times the observing bandwidth (in GHz).  For
continuum observations at a constant sampling rate, 2-bit quantization achieves
sensitivity levels very nearly approaching that of 1-bit quantization at half
the observing bandwidth \citep{thompson01}, which is limited by the hardware at
certain telescopes (\S \ref{antennas}).

Observations of total intensity (i.e., Stokes $I$) theoretically
obtain identical noise levels regardless of whether the data are taken
at full bandwidth in single-polarization mode or at half bandwidth in
dual-polarization mode, provided that circularly-polarized feeds are
used.  The hot spot models do not produce circular polarization
(Stokes $V$) but do produce large amounts of linear polarization
(Stokes $Q$ and $U$), especially on small spatial scales.
Circularly-polarized feeds measures Stokes visibilities $I \pm V$ and
therefore are identical for our models.  Dual-polarization
observations are essential if linearly-polarized feeds are used, since
the parallel-hand data are sensitive to the Stokes visibilities $I \pm
P$, where $P$ is a linear combination of Stokes $Q$ and $U$ (depending
on feed orientation).  In practice the observer will usually prefer to
observe in dual-polarization mode when available even using circular
feeds, since cross-hand correlation products provide polarimetric
information as well.  The polarimetric products are of special
interest, since they may show asymmetries not seen in total intensity
\citep[e.g.,][]{bromley01}.  However, we focus on total intensity
methods and results in this work and defer polarimetric considerations
to a future manuscript.

\begin{figure*}                                                                                                         
\resizebox{0.19\hsize}{!}{\includegraphics{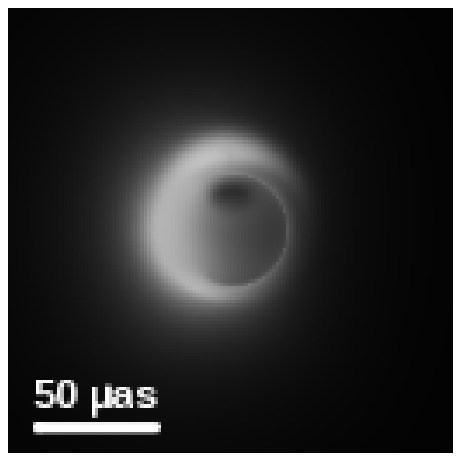}}                                                                    
\resizebox{0.19\hsize}{!}{\includegraphics{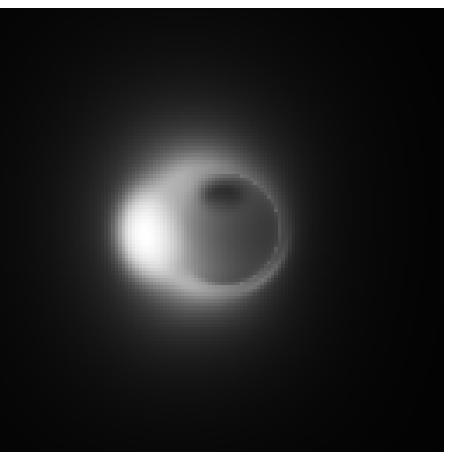}}                                                                    
\resizebox{0.19\hsize}{!}{\includegraphics{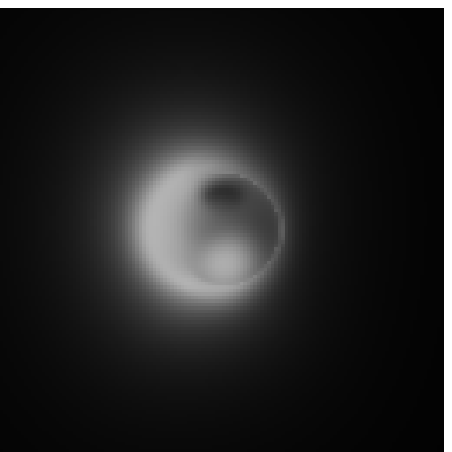}}                                                                    
\resizebox{0.19\hsize}{!}{\includegraphics{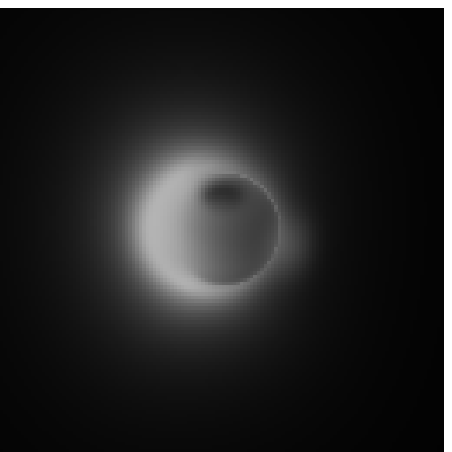}}                                                                    
\resizebox{0.19\hsize}{!}{\includegraphics{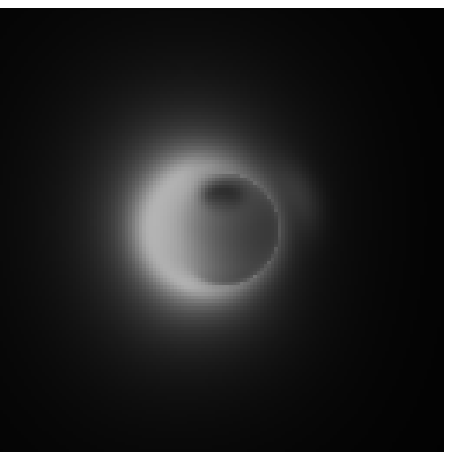}}\\                                                                  
\resizebox{0.19\hsize}{!}{\includegraphics{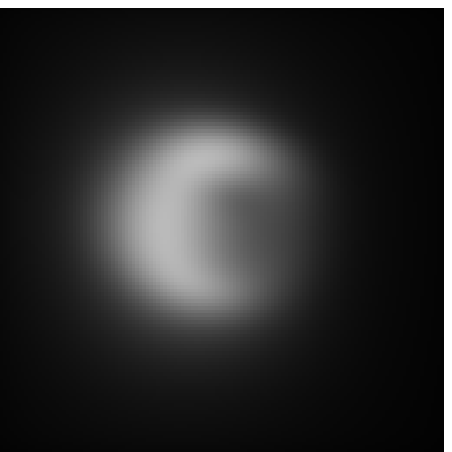}}                                                                    
\resizebox{0.19\hsize}{!}{\includegraphics{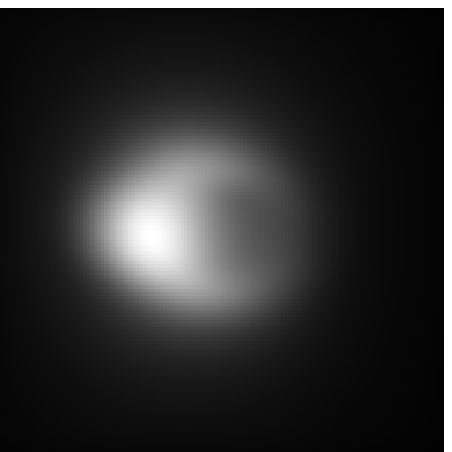}}                                                                    
\resizebox{0.19\hsize}{!}{\includegraphics{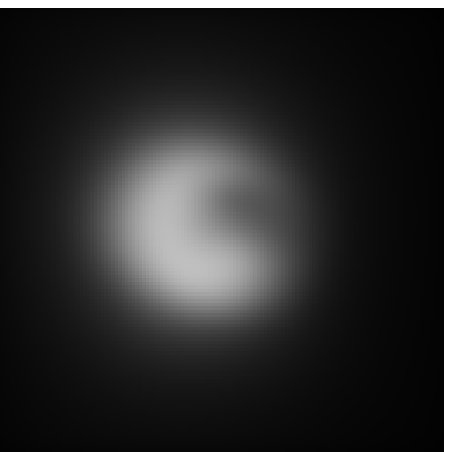}}                                                                    
\resizebox{0.19\hsize}{!}{\includegraphics{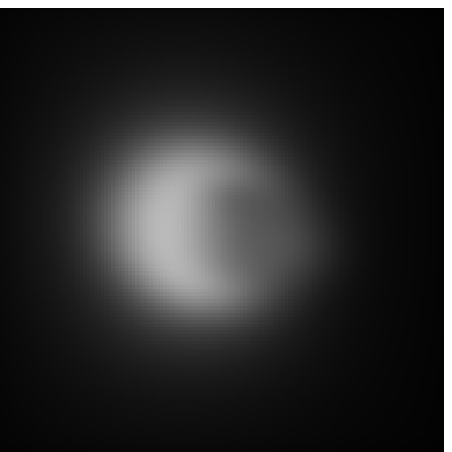}}                                                                    
\resizebox{0.19\hsize}{!}{\includegraphics{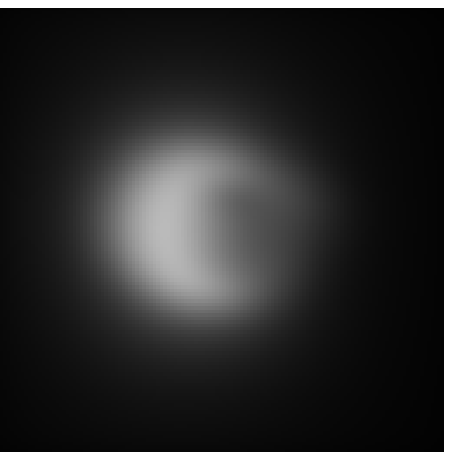}}\\                                                                  
\resizebox{0.19\hsize}{!}{\includegraphics{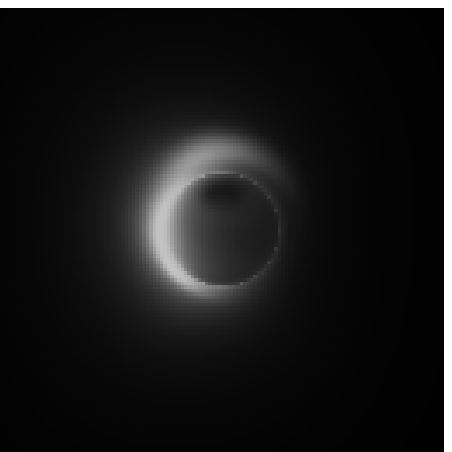}}                                                                    
\resizebox{0.19\hsize}{!}{\includegraphics{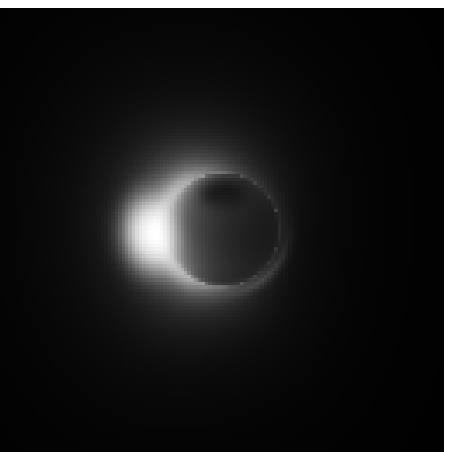}}                                                                    
\resizebox{0.19\hsize}{!}{\includegraphics{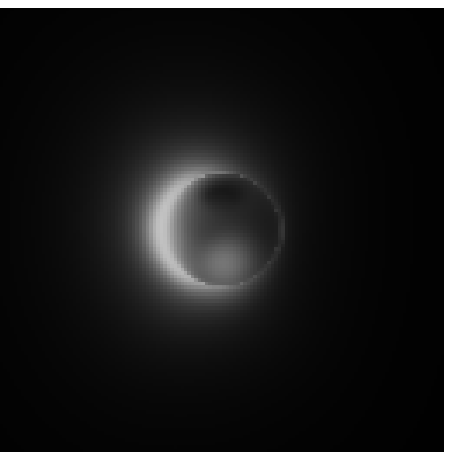}}                                                                    
\resizebox{0.19\hsize}{!}{\includegraphics{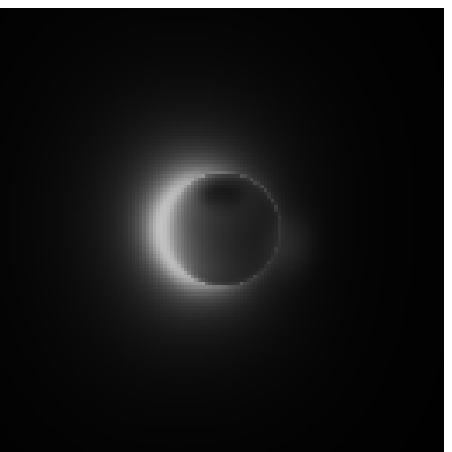}}                                                                    
\resizebox{0.19\hsize}{!}{\includegraphics{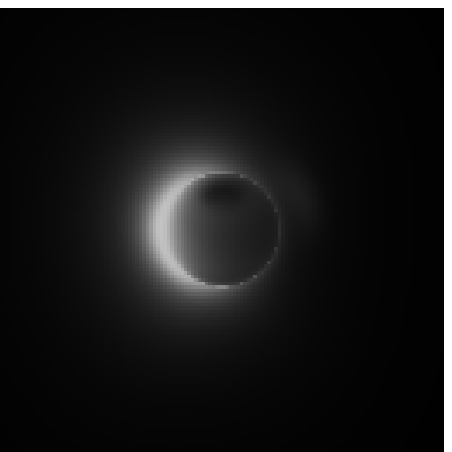}}\\                                                                  
\resizebox{0.19\hsize}{!}{\includegraphics{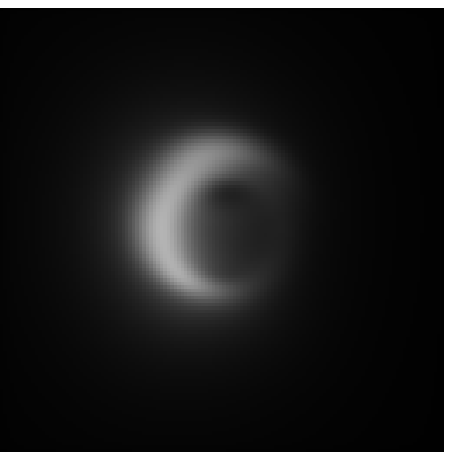}}                                                                    
\resizebox{0.19\hsize}{!}{\includegraphics{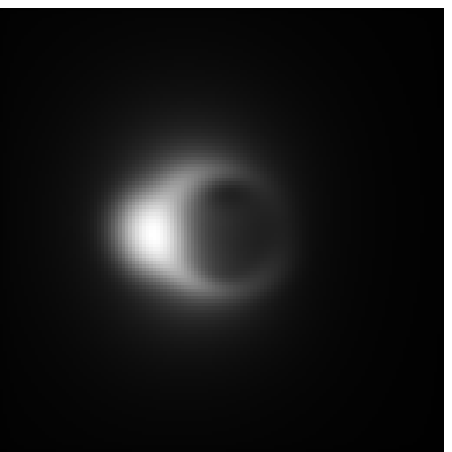}}                                                                    
\resizebox{0.19\hsize}{!}{\includegraphics{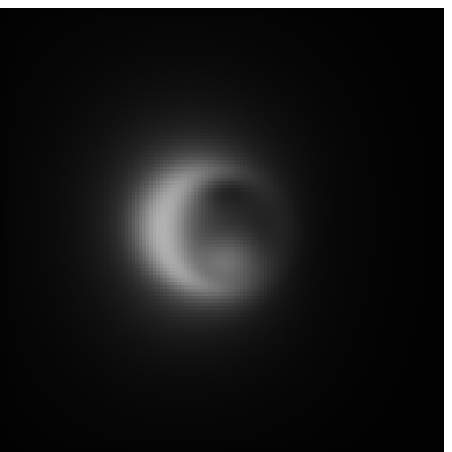}}                                                                    
\resizebox{0.19\hsize}{!}{\includegraphics{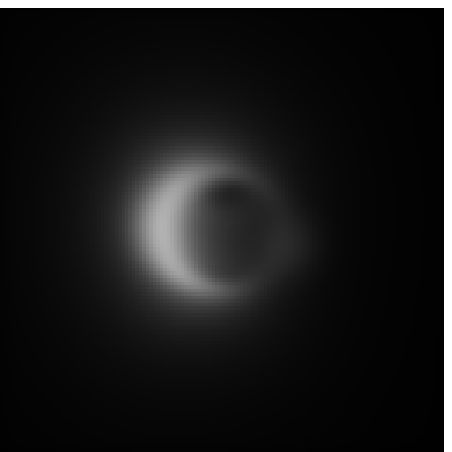}}                                                                    
\resizebox{0.19\hsize}{!}{\includegraphics{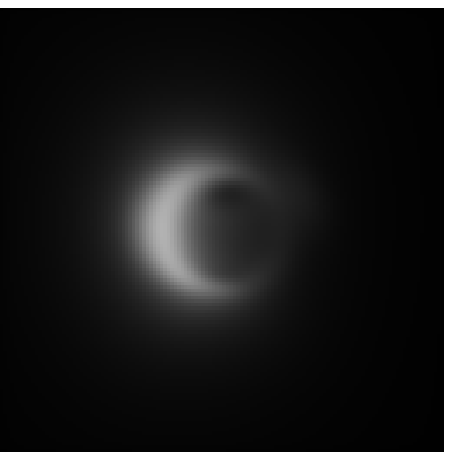}}                                                                    
\caption{Images of Model B.  Columns show orbital phases 0, 0.2, 0.4,                                                   
  0.6, and 0.8 from left to right.  From top to bottom, rows show the                                                   
  input model at 230~GHz, the same model after convolution with                                                         
  expected interstellar scattering at 230~GHz, the model at 345~GHz,                                                    
  and the same model after scattering at 345~GHz.  
\label{fig-modelb}
}                                                                                                                       
\end{figure*}

In the subsequent discussion, we consider a suite of models
parameterized by black hole spin ($a = 0$ nonrotating or $a = 0.9$
highly rotating), disk orientation, and hot spot orbital radius ($r$).
For each spin, the radius of the ISCO ($r_{ISCO} = 6 \, R_G$ for $a =
0$ and $2.32 \, R_G$ for $a = 0.9$)
and one larger radius are chosen; the larger radius for $a = 0.9$ is
chosen to have the same period as the $a = 0$, ISCO model.  Disk
models assume that the spin axes of the black hole and accretion disk
are aligned, with the major axis of the projected disk aligned
east-west except in Model C, in which the disk and hot spot are
aligned north-south instead.  Flux densities have been scaled to match
connected-element interferometric observations.
Contributions from the quiescent disk are assumed to be approximately
3~Jy, depending slightly on the specific model parameters chosen, as
listed in Table~\ref{tab-models}.  A single prograde orbiting hot spot
contributes a variable flux density component, depending on orbital
phase and the specific model, consistent with submillimeter
observations \citep{marrone07b}.  The hot
spot contributes some flux density even at the minimum in the light
curve, an effect that is most pronounced in the $i = 30$\degr\ models.
All models are convolved with the expected interstellar scattering given by
\citep{bower06}.  While these model parameters do not span the entire range of
possibilities of the Sgr A* disk system, they do show how changes in model
parameters affect the observable quantities.  Non-imaging VLBI methods are
applicable to all reasonable models for the Sgr A* system.  Figure~\ref{fig-modelb} shows 
a series of images from Model B at both 230~GHz and 345~GHz.

\begin{deluxetable}{lllrlrlll}
\tablewidth{\hsize}
\tablecaption{Model Parameters\label{tab-models}}
\tablehead{
  \colhead{} &
  \colhead{$a$} &
  \colhead{Period} &
  \colhead{$i$} &
  \colhead{PA\tablenotemark{b}} &
  \colhead{$\nu$} &
  \colhead{Disk\tablenotemark{c}} &
  \colhead{Min\tablenotemark{c}} &
  \colhead{Max\tablenotemark{c}}
\\
  \colhead{Model} &
  \colhead{($R_G$)\tablenotemark{a}} &
  \colhead{(min)} &
  \colhead{(\degr)} &
  \colhead{(\degr)} &
  \colhead{(GHz)} &
  \colhead{(Jy)} &
  \colhead{(Jy)} &
  \colhead{(Jy)}
}
\startdata
 A & 0   & 27.0 & 30 & 90 & 230 & 3.19 & 3.49 & 4.05 \\
   &     &      &    &    & 345 & 3.36 & 3.63 & 5.28 \\
 B & 0   & 27.0 & 60 & 90 & 230 & 3.03 & 3.05 & 4.03 \\
   &     &      &    &    & 345 & 2.96 & 2.99 & 4.78 \\
 C & 0   & 27.0 & 60 &  0 & 230 & 3.03 & 3.05 & 4.03 \\
   &     &      &    &    & 345 & 2.96 & 2.99 & 4.78 \\
 D & 0.9 & 27.0 & 60 & 90 & 230 & 2.98 & 2.99 & 4.05 \\
   &     &      &    &    & 345 & 2.96 & 2.97 & 4.00 \\
 E & 0.9 &  8.1 & 60 & 90 & 230 & 2.98 & 3.08 & 4.15 \\
   &     &      &    &    & 345 & 2.96 & 3.04 & 6.07 \\
 F & 0   &166.9 & 60 & 90 & 230 & 3.07 & 3.08 & 3.38 \\
   &     &      &    &    & 345 & 2.99 & 3.00 & 3.18 
\enddata
\tablenotetext{a}{Spin is given in units of the gravitational radius,
  $R_G \equiv GMc^{-2} = \onehalf\,\Rs$.}
\tablenotetext{b}{Disk major axis position angle (east of
  north).}
\tablenotetext{c}{Flux density of quiescent disk alone and
  minimum/maximum of system with orbiting hot spot.}
\end{deluxetable}

\begin{figure}
\resizebox{\hsize}{!}{\includegraphics{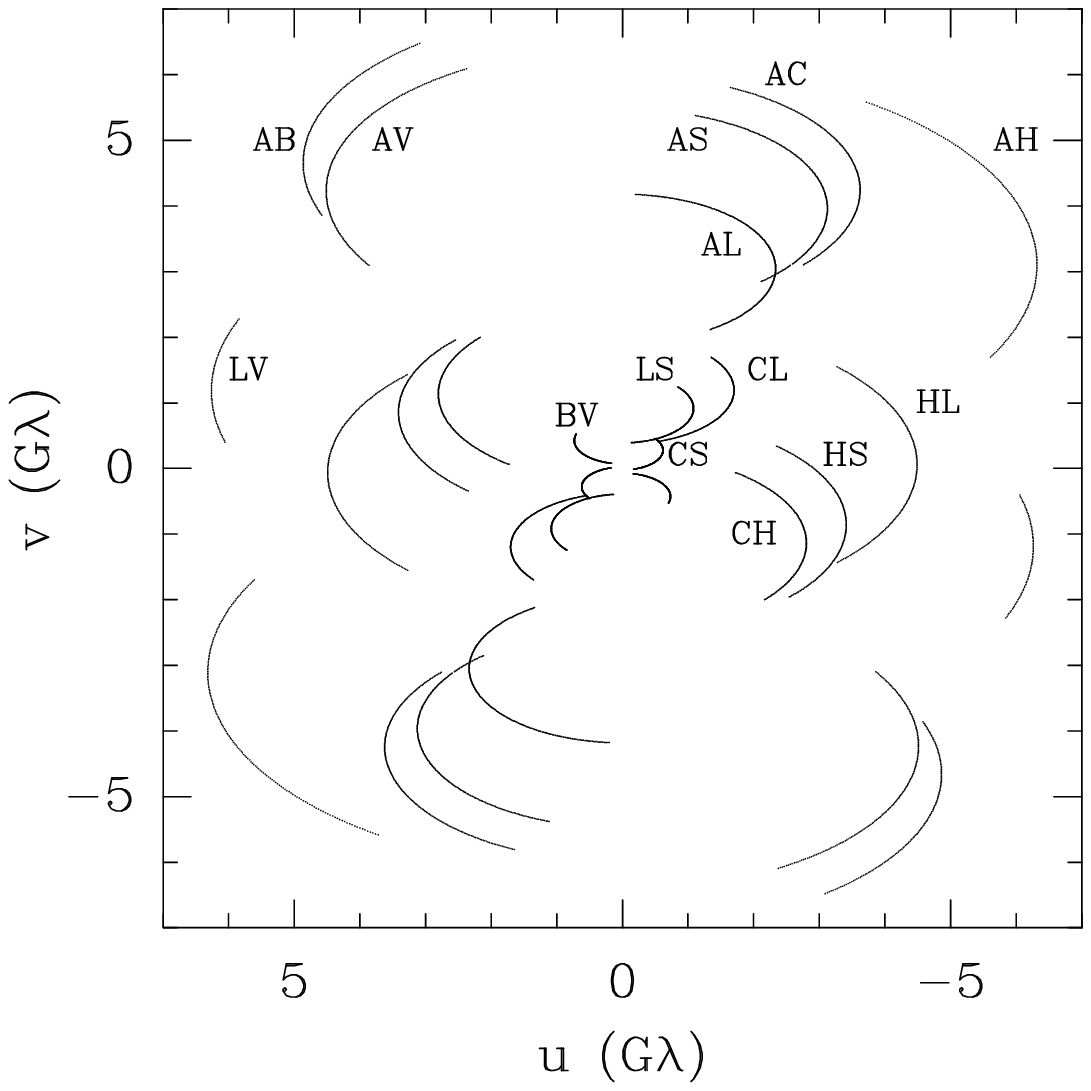}}
\caption{Possible $(u,v)$ tracks for millimeter VLBI.  Tracks are
  labelled by baseline (\emph{A}: ALMA/APEX/ASTE, \emph{B}: Plateau de
  Bure, \emph{C}: CARMA, \emph{H}: Hawaii, \emph{L}: LMT, \emph{S}:
  SMTO, \emph{V}: Pico Veleta).  Unlabelled tracks correspond to the
  baseline indicated by $(-u,-v)$.  Axes are in units of
  gigawavelengths at $\nu = 230$~GHz.
\label{fig-uvplt}}
\end{figure}

\begin{figure}
\resizebox{\hsize}{!}{\includegraphics{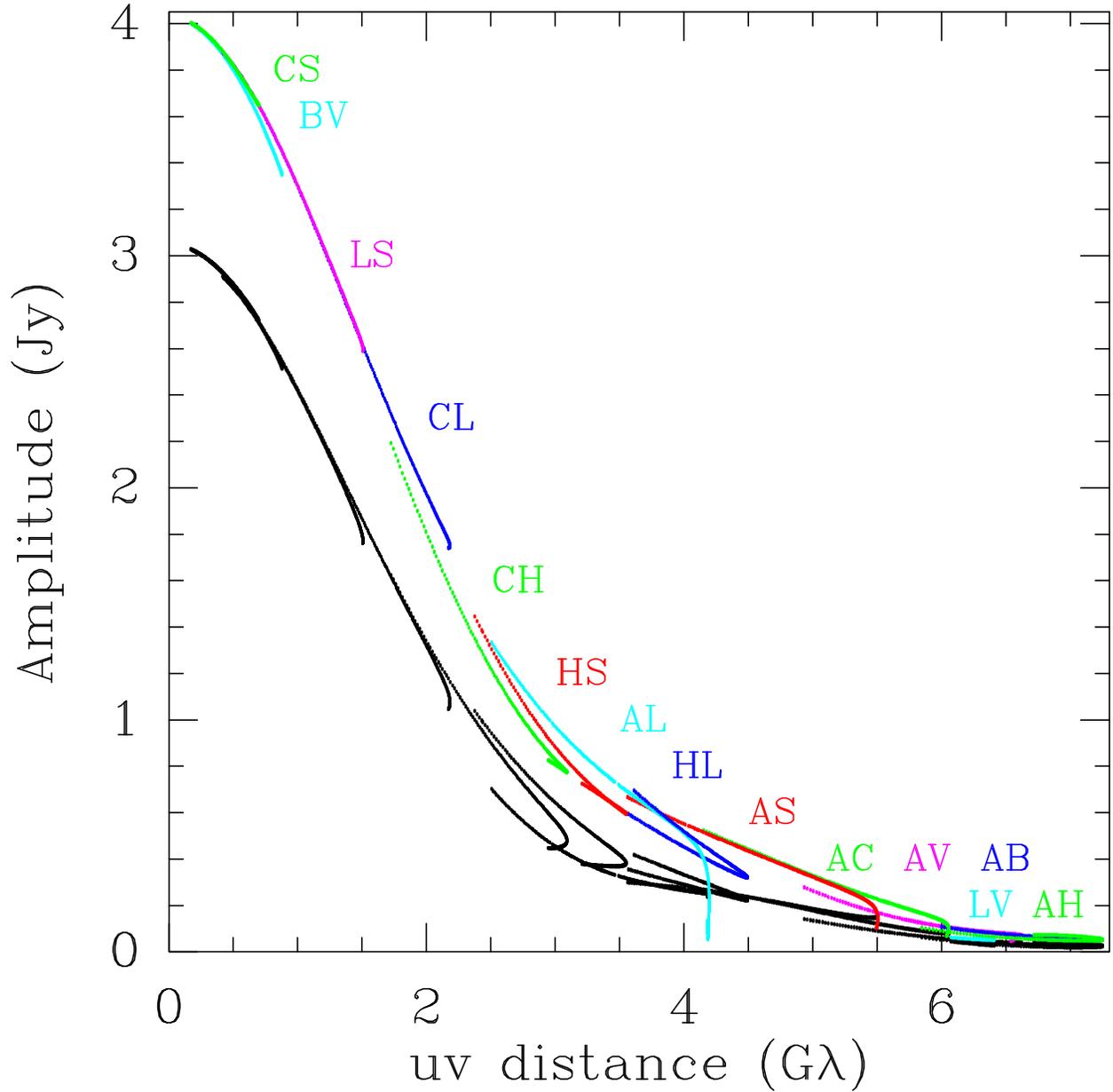}}
\caption{Plot of expected visibility amplitude as a function of
  baseline length for two frames of Model B230.  Symbols show
  the noiseless visibility amplitudes that would be obtained if the
  source structure were frozen near minimum (black) and maximum
  (color) flux in the hot spot orbit.  At maximum flux, points are
  color-coded and labelled by baseline as in Figure~\ref{fig-uvplt}.
  Visibility amplitude falls off rapidly with baseline length, and the
  fractional variability on long baselines can be significantly
  different than that seen at zero spacing. 
\label{fig-uv}}
\end{figure}

The angular resolution available to potential millimeter VLBI arrays
is well matched to potentially interesting scales for observing Sgr~A*
(Fig.~\ref{fig-uvplt}).  The longest baselines, Hawaii-Chile and
PdB-Chile, provide fringe spacings of 30--35~$\mu$as at 230~GHz and
19--23~$\mu$as at 345~GHz, slightly larger than the expected
interstellar scattering and only several times $\Rs$.  Our
models do not produce much detectable signal on these small angular
scales (Fig.~\ref{fig-uv}), but it is possible that smaller hot spots
or disk instabilities (not modelled) will produce greater amplitudes
on small angular scales.  The shortest baselines, SMTO-CARMA and
PV-PdB, provide fringe spacings of 230--1000~$\mu$as at 230~GHz and
160--700~$\mu$as at 345~GHz.

\section{Results}
\label{results}

\subsection{Closure Phases and Amplitudes}

\begin{figure*}
\resizebox{\hsize}{!}{\rotatebox{-90}{\includegraphics[]{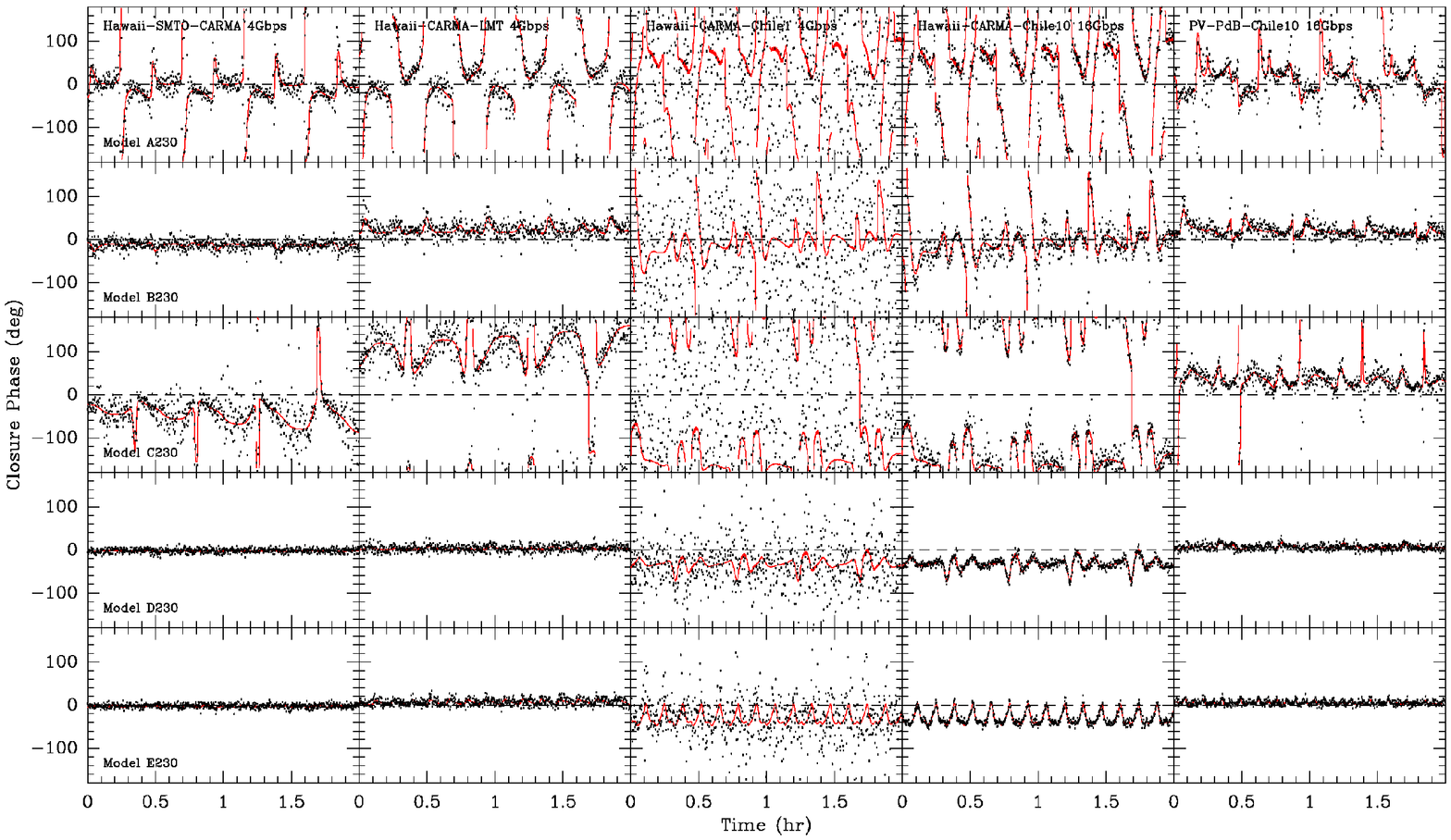}}}
\caption{Closure phases on selected triangles at 230~GHz.  The solid
  line (red in the online edition) shows the predicted closure phase
  in the absence of noise.  Each point indicates 10~s of
  coherently-integrated data.  The fourth column shows the same
  triangle as the third column but with a higher data rate and
  substitution of Chile~10 for Chile~1.  The same 2-hour period,
  corresponding to 4.5 periods (14.8 periods for Model E) is
  shown in all panels excepting PV-PdB-Chile~10.
\label{fig-phase230}}
\end{figure*}

Figure~\ref{fig-phase230} shows predicted closure phases at 230~GHz on
selected triangles.  The closure phase responses are highly dependent
on the physical parameters of the Sgr~A* system, but several patterns
emerge.  Data from the smallest triangles (i.e., those composed of
only North American telescopes) achieve a high SNR even at the present
maximum recording rate of 4~Gbit\,s$^{-1}$.  Models with a greater
north-south extent (e.g., Models A and C), produce a larger closure
phase signature on the small triangles.  The SNR on all triangles to
Chile~1 is low owing to the small flux on small angular scales.
Higher bit rates and the substitution of phased ALMA for Chile~1
greatly increase detectability, as shown by the third and fourth
columns, which correspond to the expected array and recording
capabilities in the next few years.  Net offsets from zero closure
phase on some triangles result from apparent asymmetric structure in
the disk caused by lensing and opacity effects.

\begin{figure*}
\resizebox{\hsize}{!}{\rotatebox{-90}{\includegraphics[]{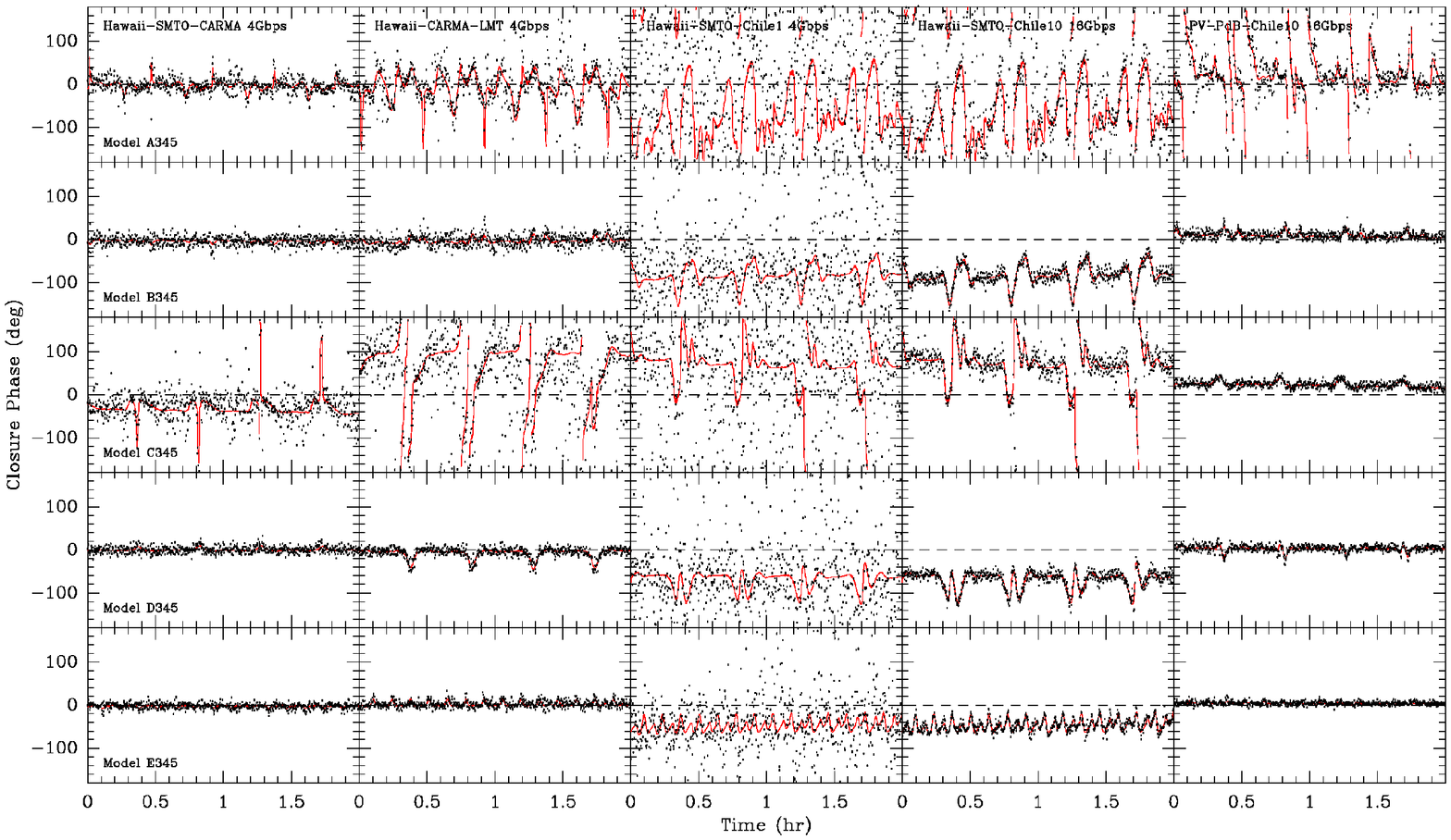}}}
\caption{Closure phases on selected triangles at 345~GHz.  See
  Figure~\ref{fig-phase230} for details.
\label{fig-phase345}}
\end{figure*}

Figure~\ref{fig-phase345} shows modelled closure phases at 345~GHz.
The only available triangle in the near future at 345~GHz will be
Hawaii-SMTO-Chile~1, which produces rather low SNR closure phases at
4~Gbit\,s$^{-1}$.  Of the remaining telescopes, both CARMA and Plateau
de Bure plan for 345~GHz receivers in the future.  In addition to
these, we show triangles including the LMT and Pico Veleta, neither of
which has planned 345~GHz capability, to demonstrate what might be
seen in the event of future upgrades to those facilities.  At both
230~GHz and 345~GHz, closure phases on several baseline triangles show
clear evidence for periodicity associated with hot-spot orbits.

\begin{figure*}
\resizebox{\hsize}{!}{\rotatebox{-90}{\includegraphics[]{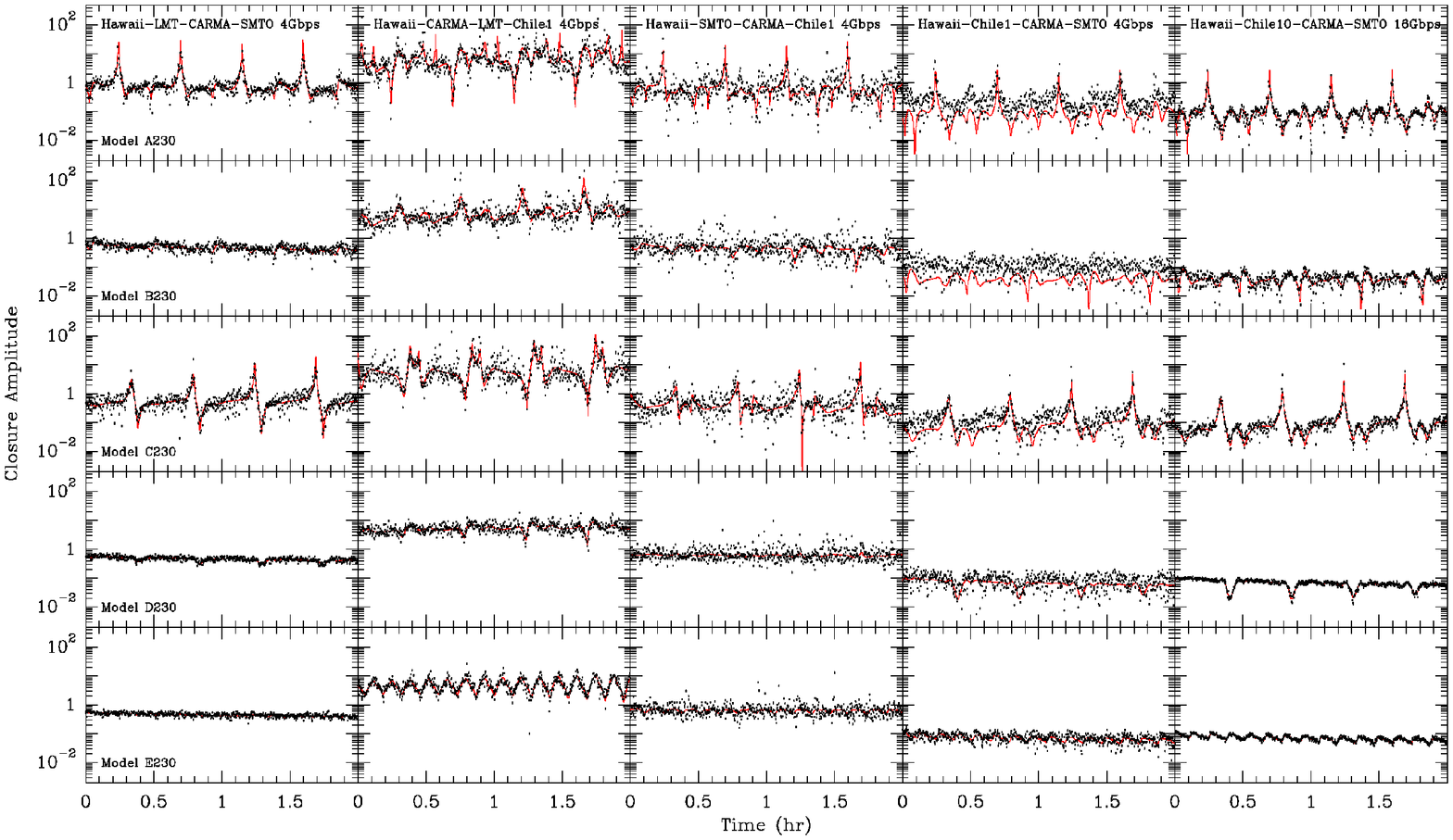}}}
\caption{Closure amplitudes on selected quadrangles at 230~GHz.  See
  Figure~\ref{fig-phase230} for details.
\label{fig-amp230}}
\end{figure*}

\begin{figure*}
\resizebox{\hsize}{!}{\rotatebox{-90}{\includegraphics[]{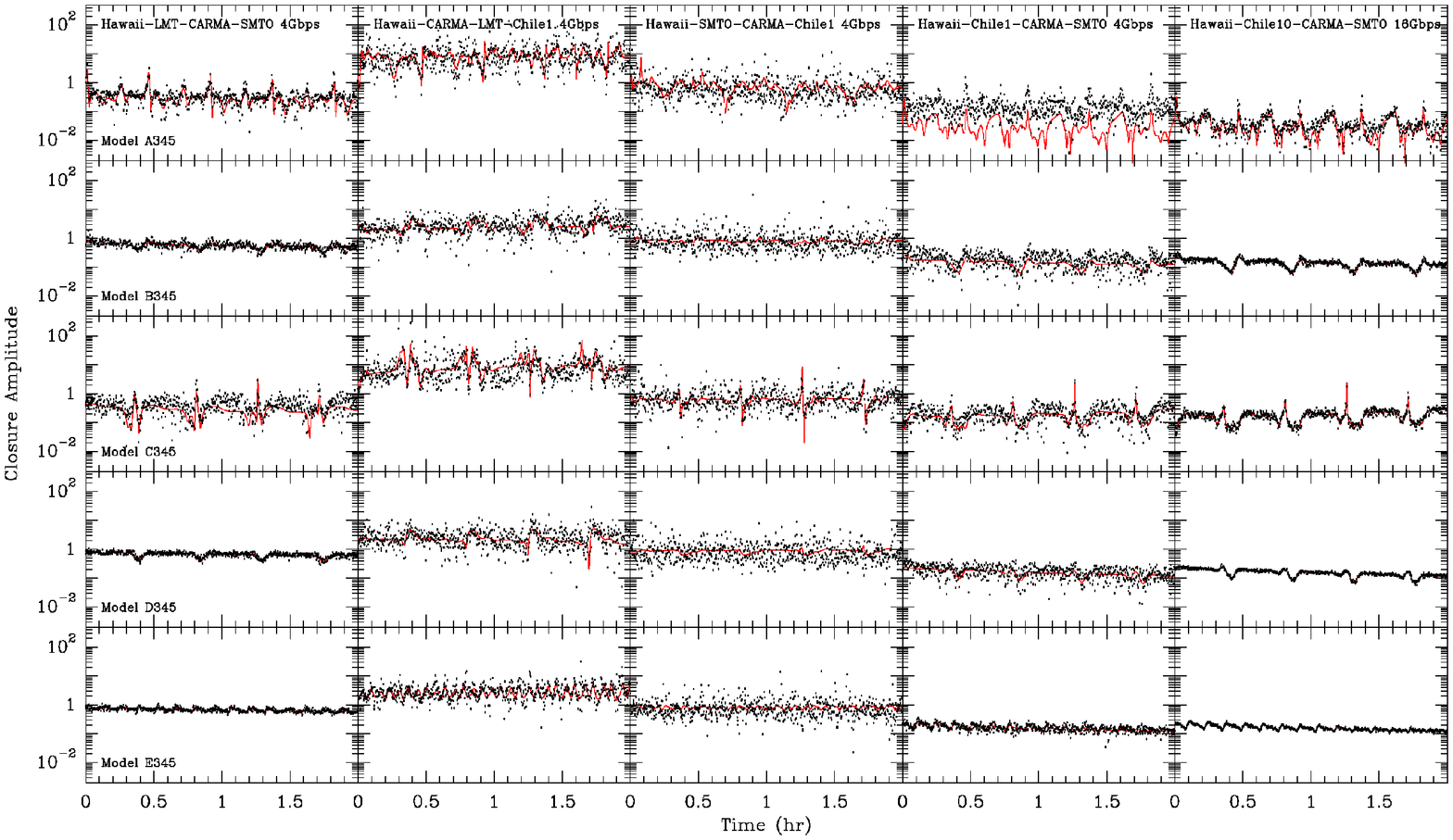}}}
\caption{Closure amplitudes on selected quadrangles at 345~GHz.  See
  Figure~\ref{fig-phase230} for details.
\label{fig-amp345}}
\end{figure*}

Four telescopes are required to obtain a closure amplitude, which
effectively means that closure amplitudes can only be measured on
Western hemisphere arrays.  Figures~\ref{fig-amp230} and
\ref{fig-amp345} show closure amplitudes on selected quadrangles at
230 and 345~GHz, respectively.  Periodicity associated with the
orbiting hot-spot is evident on most quadrangles.  In some models,
closure amplitudes including the Hawaii-Chile~1 baseline show bias (as
described in \S \ref{techniques}) due to the low visibility amplitude
on this baseline.  The substitution of phased ALMA for Chile~1
combined with higher recording rates suffice to clearly detect Sgr~A*
on this baseline, resulting in unbiased closure amplitudes.  It is
possible (and necessary for model fitting) to de-bias closure
amplitudes by correcting the visibility amplitudes for the expected
noise levels, but the procedure can be difficult when the SNR on a
baseline is low \citep[for more details, see][]{trotter98}.  In any
case, the presence of bias does not hinder detection of periodicity.

\subsection{Autocorrelation Functions}
\label{ACF}

Signatures of time periodic structure associated with orbiting
hot-spots can be derived from autocorrelations of closure quantity
time series.  The autocorrelation function of a time series of $n$
closure amplitudes $A$ on a quadrangle of telescopes is given by
\begin{equation}
\mathrm{ACF}_A(k) \equiv \frac{1}{(n-k)\,\sigma^2}
\sum\limits_{i=1}^{n-k}\left[\left(\log A_i-\mu\right)\left(\log
A_{i+k}-\mu\right)\right],
\end{equation}
where $\mu$ and $\sigma^2$ are the mean and variance of the
distribution of the logarithm of the closure amplitudes, respectively.
The logarithm is used in preference to the closure amplitude itself
due to the tendency for the closure amplitude to obtain both very
small (near 0) and very large values.  We also define a variant of the
autocorrelation function for closure phases, $\phi$:
\begin{equation}
\mathrm{ACF}_\phi(k) \equiv \frac{1}{n-k}
\sum\limits_{i=1}^{n-k}\cos\left(\phi_i-\phi_{i+k}\right).
\end{equation}
This definition has the advantage that it handles phase-wrap ambiguity
gracefully, since $\cos$ is a periodic function.  The normalizations
are such that $\mathrm{ACF}(k) = 1$ when $k$ is an integer multiple of
the period for a noiseless periodic function.  In practice, the peaks
of an ACF (other than the trivial peak at $k = 0$, where
$\mathrm{ACF}(0) = 1$ by definition) of closure phases or amplitudes
fall off with lag rather than returning precisely to unity, because
the projected baseline geometries change with Earth rotation.  In the
case of long periodicities ($\gtrsim 2$~hr), this effect can be large
enough to obscure any periodicity at all on some
triangles/quadrangles.  While the duration of some SgrA* flare events
exceed this time interval, there are claims of modulation within some
NIR flares with characteristic time scales of $\sim17$~min
\citep{genzel03}.  From an observational perspective, periodicity is
not convincingly detected until at least two full periods (preferably
more) have been observed.  The longest mutual visibility appears on
the SMTO-LMT-Chile triangle, which can see Sgr~A* for less than 7 full
hours.  Triangles including Hawaii or Europe have significantly
smaller windows of mutual visibility.  The distribution of millimeter
telescopes is not optimal for detecting slow periodic variability in
Sgr~A*.

\begin{figure*}
\resizebox{\hsize}{!}{\rotatebox{-90}{\includegraphics{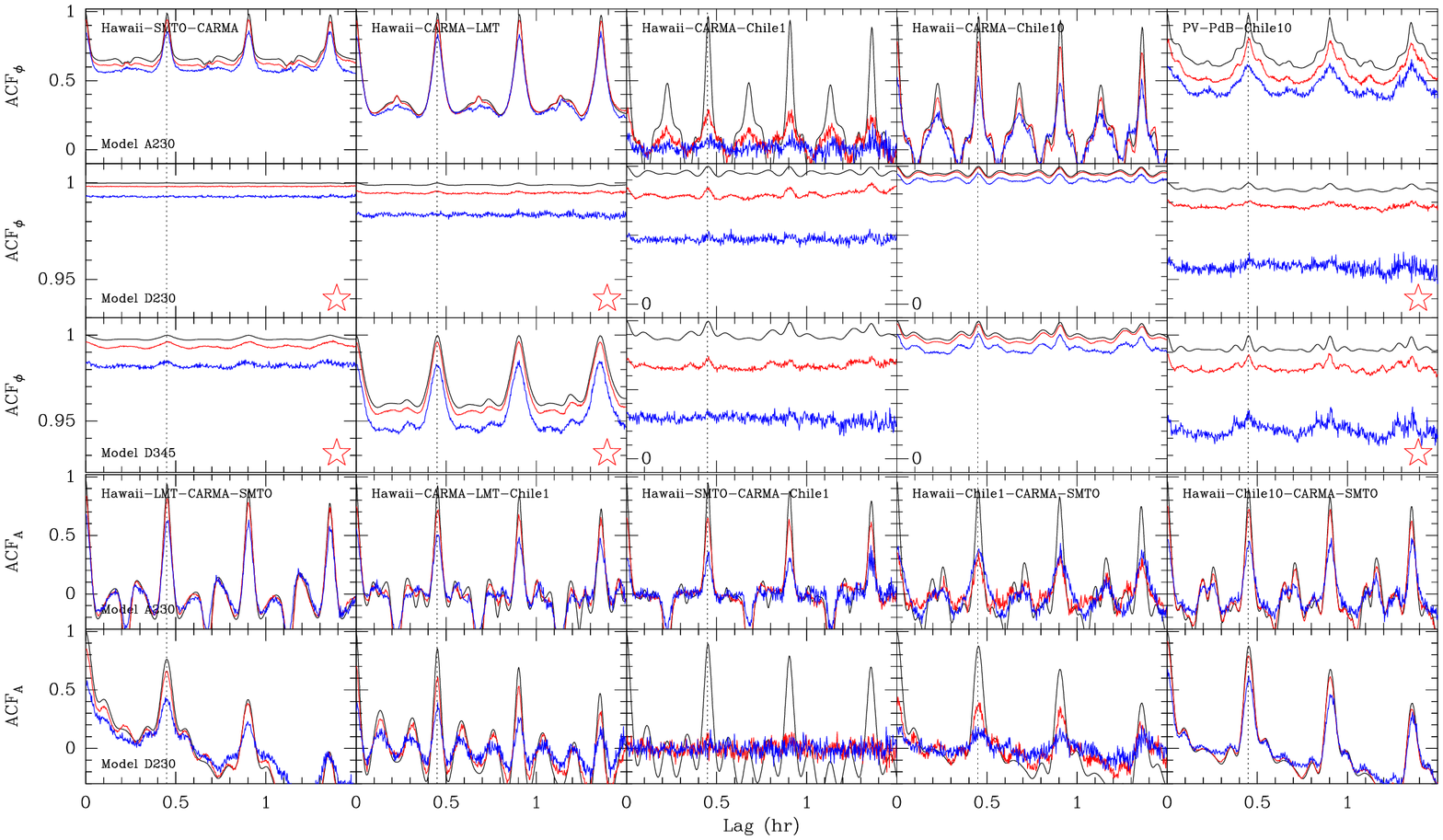}}}
\caption{Autocorrelation function plots of selected triangles and
  quadrangles.  Representative models are shown.  Black, red, and blue
  lines indicate ACFs for noiseless data, 16~Gbit\,s$^{-1}$, and
  4~Gbit\,s$^{-1}$, respectively.  Panels marked with a red star show
  $\mathrm{ACF}_\phi$ on a different ordinate scale, as indicated to
  the left in that row; other panels in the same row are on the scale
  show in the upper left panel.  The dotted line shows the period of
  the hot spot orbit.
\label{fig-acfs}}
\end{figure*}

The largest nontrivial peak in the ACF indicates the period of the hot
spot orbit, as indicated in Figure~\ref{fig-acfs}, excluding the slow
periodic case.  The ACF correctly identifies the period in all models
at all recording rates provided that the average SNR $\gtrsim 1$ and
that the triangle contains at least one baseline long enough to be
sensitive to the changing source structure.  On the small triangles,
$\mathrm{ACF}_\phi(k)$ shows little variation as a function of lag
$k$, since the phase autocorrelation function of a nearly-constant
function is itself nearly constant.  However, small peaks are visible
in the ACF on the Hawaii-SMTO-CARMA and Hawaii-CARMA-LMT triangles in
Figure~\ref{fig-acfs}.  The large SNR on the small triangles, due to
the fact that the shorter baselines do not resolve out much of the
source flux density, permits detectability for reasonable bandwidths.

In addition to the peak at the period, the ACF has peaks at integer
multiples of the period.  In the weak-detection limit, this may be an
important discriminator indicating that the detected period is real.
While purely random noise may produce peaks in the ACF, there is no
reason why it should produce periodic peaks.  At the other end of time
scales, intraperiod sub-peaks in the ACF may be produced depending on
the details of the source structure and array geometry.  Excluding
long-period orbits, it is usually clear which peak indicates the
orbital period of the hot spot, since the sub-peaks are of smaller
amplitude.  Nevertheless, it is possible that pathological cases exist
in which the observer may misidentify the orbital period.  For
instance, two identical hot spots at the same radius but separated by
180\degr\ in azimuth would produce strong peaks in the ACF at both
integer and half-integer multiples of the period, which would lead to
a conclusion that the orbital period is half of its true value.  A
similar ambiguity could arise if the observed variability is produced,
for instance, by the rotation of a pattern produced by a two-armed
Rossby wave instability \citep[e.g.,][]{falanga07}.

The key point about the ACF plots is that for the same model,
\emph{all triangles and quadrangles indicate the same period}.  It
will be important to observe with as many telescopes as possible
simultaneously, since the extra information provided by the additional
telescopes may be important for detecting or confirming
marginally-detected variability.  Three antennas provide a single
closure phase, for a total of 1 ACF; four antennas provide three
independent closure phases and two independent closure amplitudes, for
a total of 5 independent ACFs; and five antennas provide six closure
phases and five closure amplitudes, for a total of 11 independent
ACFs.  Even if the source is not detected on one long baseline,
resulting in a biased closure amplitude, the ACF may still clearly
indicate periodicity (e.g., the Hawaii-Chile~1-CARMA-SMTO panel for
Model A230 in Fig.~\ref{fig-acfs}).

\subsection{Detecting Periodicity with Likely Arrays}
\label{results-arrays}

We consider potential observing arrays that may be employed in
observations of Sgr~A*.  The minimum number of telescopes required to
produce a closure quantity is 3, and the maximum number of telescopes
located at substantially different sites with mutual visibility of
Sgr~A* is 5.  At 230~GHz, we consider western hemisphere arrays
consisting of the US triangle; US and Chile-1; US, Chile-1, and the
LMT; and the same array with Chile-10 instead.  We also consider
European-Chile arrays.  Since PV is not expected to have 345~GHz
capability in the near future, we expect that Chile-10 will exist
before 345~GHz observations are possible on the Europe-Chile triangle.
Likewise, we expect that western hemisphere arrays will include
Chile-10 before the LMT is available at 345~GHz, since 345~GHz
capability is not currently planned at the LMT.

For each array, we generate $\mathrm{ACF}_\phi$ for each triangle of
three antennas and $\mathrm{ACF}_A$ for each quadrangle of four
antennas.  The ACFs can be combined to produce a single combined ACF
of the entire array.  For optimum detectability, it is necessary to
weight the individual ACFs by the square of their effective
$\mathrm{SNR}_\mathrm{ACF} \equiv X/\xi$, where $X$ and $\xi$ quantify
the effective signal and noise in the ACF, respectively.  For
definiteness, we take $\xi$ as the rms deviation in the ACF after a
5-channel boxcar-smoothed SNR has been subtracted.  We define $X
\equiv \max(\mathrm{ACF}(k)) - \max(\min(\mathrm{ACF}(k)),0), k \neq
0$ as a measure of the range of the ACF.  The quantity $X$ reflects
the contrast of the ACF and is especially critical for proper
weighting of the phase ACFs, since on many triangles
$\mathrm{ACF}_\phi$ is approximately constant though with small
periodic peaks (Fig.~\ref{fig-acfs}).  Effectively, $X/\xi$ quantifies
the significance of a peak in the ACF.  In practice, $X$ is frequently
small for $\mathrm{ACF}_\phi$ on certain triangles, although the very
high SNR of the closure phases on the small SMTO-CARMA-LMT and
Hawaii-SMTO-CARMA triangles (which do not resolve out much of the
total flux) often compensates by producing a very clean ACF (i.e.,
$\xi$ is small as well).  In order to identify the period in an
algorithmic way, we created a ``folded'' version of the composite
array ACF, defined as
\begin{equation}
\mathrm{Fold}(k) \equiv \frac{1}{\lfloor
  m/k\rfloor}\sum\limits_{i=1}^{\lfloor m/k\rfloor}
\mathrm{ACF}(i\cdot k),
\end{equation}
where $m$ is the number of points in the ACF.  Folding the ACF
effectively suppresses the trivial peak at zero lag.  In the absence
of noise, the peak of $\mathrm{Fold}(k)$ is the period.

In order to test for the significance of periodicity detection, we ran
Monte Carlo simulations for each model and array at bit rates from 1
to 32~Gbit\,s$^{-1}$ by powers of two.  We ran 10000 simulations of
the data with different noise instantiations and obtained the value of
$k$ maximizing $\mathrm{Fold}(k)$.  The distribution of
$\mathrm{Fold}(k)$ provides information on the probability of false
detection of periodicity.  The results are summarized in
Table~\ref{tab-results}, which lists the number of trials for which
$\max(\mathrm{Fold}(k))$ misidentifies the period, rounded to the
nearest 10~s, by more than one minute.  Four and a half orbital
periods of simulated data were used, corresponding to 2~hr for Models
A-D and 37~min for Model E.

\clearpage \begin{deluxetable}{lrlrrrrrr}
\tabletypesize{\scriptsize}
\tablewidth{\hsize}
\tablecaption{Probabilities of False Periodicity Detection\label{tab-results}}
\tablehead{
  \colhead{} &
  \colhead{Period\tablenotemark{a}} &
  \colhead{} &
  \colhead{$P_{32}$} &
  \colhead{$P_{16}$} &
  \colhead{$P_{8}$} &
  \colhead{$P_{4}$} &
  \colhead{$P_{2}$} &
  \colhead{$P_{1}$} \\
%
  \colhead{Model} &
  \colhead{(s)} &
  \colhead{Array} &
  \colhead{($\times 10^{-4}$)} &
  \colhead{($\times 10^{-4}$)} &
  \colhead{($\times 10^{-4}$)} &
  \colhead{($\times 10^{-4}$)} &
  \colhead{($\times 10^{-4}$)} &
  \colhead{($\times 10^{-4}$)}
}
\startdata
\multicolumn{9}{c}{230 GHz} \\
\tableline \\
A   & 1620 & JCMT, SMTO, CARMA-1                &    0 &    0 &    0 &    5 &  397 & 2199 \\
    &      & JCMT, SMTO, CARMA-1, Chile-1       &    0 &    0 &    0 &    0 &   27 &  978 \\
    &      & Hawaii, SMTO, CARMA                &    0 &    0 &    0 &    0 &    1 &  134 \\
    &      & Hawaii, SMTO, CARMA, Chile-1       &    0 &    0 &    0 &    0 &    0 &    0 \\
    &      & PV, PdB, Chile-1                   &    0 &   51 &  655 & 2516 & 5542 & 8093 \\
    &      & PV, PdB, Chile-10                  &    0 &    0 &    0 &    0 &    1 &  134 \\
B   & 1620 & JCMT, SMTO, CARMA-1                &  157 & 2216 & 6653 & 9126 & 9666 & 9566 \\
    &      & JCMT, SMTO, CARMA-1, Chile-1       &    0 &   12 &  111 &  517 & 2356 & 5696 \\
    &      & Hawaii, SMTO, CARMA                &    0 &    0 &   94 & 1827 & 6231 & 8981 \\
    &      & Hawaii, SMTO, CARMA, Chile-1       &    0 &    0 &    2 &   80 &  535 & 2119 \\
    &      & Hawaii, SMTO, CARMA, LMT, Chile-1  &    0 &    0 &    0 &    0 &    0 &    0 \\
    &      & PV, PdB, Chile-1                   &   97 &  548 & 1138 & 2473 & 5003 & 7545 \\
    &      & PV, PdB, Chile-10                  &    0 &    0 &    0 &    6 &  185 &  724 \\
C   & 1620 & JCMT, SMTO, CARMA-1                &    0 &    0 &    2 &   41 &  316 & 1426 \\
    &      & JCMT, SMTO, CARMA-1, Chile-1       &    0 &    0 &    0 &    0 &    5 &  369 \\
    &      & Hawaii, SMTO, CARMA                &    0 &    0 &    0 &    0 &    0 &   21 \\
    &      & Hawaii, SMTO, CARMA, Chile-1       &    0 &    0 &    0 &    0 &    0 &    0 \\
    &      & PV, PdB, Chile-1                   &  180 & 1993 & 4694 & 6943 & 8357 & 9183 \\
    &      & PV, PdB, Chile-10                  &    0 &    0 &    0 &   12 &  437 & 2759 \\
D   & 1620 & JCMT, SMTO, CARMA-1                & 9195 & 9679 & 9809 & 9854 & 9871 & 9883 \\
    &      & JCMT, SMTO, CARMA-1, Chile-1       &    0 &    0 &    3 &  553 & 4435 & 7702 \\
    &      & Hawaii, SMTO, CARMA                & 3689 & 7207 & 8981 & 9556 & 9770 & 9825 \\
    &      & Hawaii, SMTO, CARMA, Chile-1       &    0 &    0 &    0 &    3 &  754 & 5584 \\
    &      & Hawaii, SMTO, CARMA, LMT, Chile-1  &    0 &    0 &    0 &    0 &    0 &    0 \\
    &      & PV, PdB, Chile-1                   & 6477 & 8667 & 9337 & 9562 & 9561 & 9593 \\
    &      & PV, PdB, Chile-10                  &    0 &    1 &  349 & 3309 & 7389 & 8983 \\
E   &  490 & JCMT, SMTO, CARMA-1                & 2318 & 4953 & 6730 & 7557 & 7873 & 8098 \\
    &      & JCMT, SMTO, CARMA-1, Chile-1       &    0 &    0 &   70 &  991 & 3208 & 5240 \\
    &      & Hawaii, SMTO, CARMA                &    1 &  204 & 1753 & 4394 & 6385 & 7390 \\
    &      & Hawaii, SMTO, CARMA, Chile-1       &    0 &    0 &    0 &   34 &  772 & 3238 \\
    &      & Hawaii, SMTO, CARMA, LMT, Chile-1  &    0 &    0 &    0 &    0 &    0 &    7 \\
    &      & Hawaii, SMTO, CARMA, LMT, Chile-10 &    0 &    0 &    0 &    0 &    0 &    0 \\
    &      & PV, PdB, Chile-1                   & 5987 & 6943 & 7461 & 7300 & 7037 & 7184 \\
    &      & PV, PdB, Chile-10                  &   57 &  778 & 2564 & 4750 & 6454 & 7211 \\
\tableline
\multicolumn{9}{c}{345 GHz} \\
\tableline \\
A   & 1620 & Hawaii, SMTO, Chile-1              &    0 &   20 & 1488 & 6582 & 8890 & 9505 \\
    &      & Hawaii, SMTO, CARMA                &    0 &    0 &    0 &   11 &  395 & 2588 \\
    &      & Hawaii, SMTO, CARMA, Chile-1       &    0 &    0 &    0 &    0 &   90 & 1062 \\
    &      & Hawaii, SMTO, CARMA, Chile-10      &    0 &    0 &    0 &    0 &    0 &    0 \\
    &      & PV, PdB, Chile-10                  &    0 &    0 &    0 &    0 &    0 &   26 \\
B   & 1620 & Hawaii, SMTO, Chile-1              &    0 &    0 &   45 & 2276 & 7064 & 9076 \\
    &      & Hawaii, SMTO, CARMA                &    4 &  424 & 4718 & 8504 & 9541 & 9785 \\
    &      & Hawaii, SMTO, CARMA, Chile-1       &    0 &    0 &    0 &  126 & 2676 & 7086 \\
    &      & Hawaii, SMTO, CARMA, Chile-10      &    0 &    0 &    0 &    0 &    0 &    0 \\
    &      & PV, PdB, Chile-10                  &    0 &    0 &    5 &  767 & 5066 & 8194 \\
C   & 1620 & Hawaii, SMTO, Chile-1              &    0 &    0 &    1 &  366 & 4675 & 8560 \\
    &      & Hawaii, SMTO, CARMA                &    0 &    0 &    0 &    1 &   71 & 1210 \\
    &      & Hawaii, SMTO, CARMA, Chile-1       &    0 &    0 &    0 &    0 &    2 &  161 \\
    &      & Hawaii, SMTO, CARMA, Chile-10      &    0 &    0 &    0 &    0 &    0 &    0 \\
    &      & PV, PdB, Chile-10                  &    0 &    0 &    0 &   43 & 2154 & 7393 \\
D   & 1620 & Hawaii, SMTO, Chile-1              &    0 &    0 &  137 & 3147 & 7632 & 9110 \\
    &      & Hawaii, SMTO, CARMA                &    0 &    0 &    8 &  582 & 3582 & 6901 \\
    &      & Hawaii, SMTO, CARMA, Chile-1       &    0 &    0 &    0 &   19 & 1933 & 6664 \\
    &      & Hawaii, SMTO, CARMA, Chile-10      &    0 &    0 &    0 &    0 &    0 &    0 \\
    &      & PV, PdB, Chile-10                  &    0 &    0 &   73 & 2183 & 6871 & 8590 \\
E   &  490 & Hawaii, SMTO, Chile-1              &  105 & 1057 & 2453 & 3554 & 4307 & 5617 \\
    &      & Hawaii, SMTO, CARMA                &    0 &    3 &  192 & 1705 & 4099 & 5838 \\
    &      & Hawaii, SMTO, CARMA, Chile-1       &    0 &    0 &   61 &  924 & 2112 & 3137 \\
    &      & Hawaii, SMTO, CARMA, Chile-10      &    0 &    0 &    0 &    0 &    0 &    5 \\
    &      & PV, PdB, Chile-10                  &   34 &  890 & 3183 & 5088 & 6093 & 6466   
\enddata
\tablecomments{Right columns indicate probabilities of false period
  identifications (more than 60~sec from true period) in 10000 trial
  runs at subscripted bit rate in Gbit\,s$^{-1}$.  Arrays are not
  listed when a listed proper subset of the array produces no false
  detections at 1~Gbit\,s$^{-1}$ (e.g.,
  Hawaii,SMTO,CARMA,[LMT],Chile-10 in Model A at 230~GHz).}
%
%
\tablenotetext{a}{Rounded to nearest 10~s.}
\end{deluxetable}
 \clearpage

It is clear that the inclusion of a fourth or fifth telescope in the western
hemisphere array produces a large improvement in periodicity detectability.
This is due both to the much larger number of closure quantities that can be
averaged together to produce a detection and also to the fact that a 4- or
5-element array will necessarily probe a larger range of spatial scales than
possible 3-element arrays, which is important because it is not clear a priori
which triangle will be best matched to the angular resolution of the variable
emission in Sgr~A*.  Additional bandwidth is important as well, but less so
than additional telescopes in the observing array.  The Hawaii-SMTO-CARMA
array, for example, is insensitive to periodic structure in several models,
even at high recording rates.  But inclusion of a Chilean telescope, and/or the
LMT, produces a robust detection of periodicity at modest recording rates.  It
is especially worth noting that a bit rate of 8~Gbit\,s$^{-1}$, corresponding
to 2~GHz total bandwidth (e.g., 1~GHz in each of two orthogonal polarizations)
is sufficient to produce clear periodicity detections (with error rate $< 1
\times 10^{-4}$) in nearly all models provided that at least four telescopes
are used in the array.  For the Europe-Chile triangle, where no clear fourth
telescope is presently available, the use of high bit rates and phased ALMA
will be critical for periodicity detection.

Array options at 345~GHz are relatively limited, since neither the LMT
nor PV have planned 345~GHz capability.  We include the Europe-Chile
triangle to show what might be expected if PV is ever upgraded to
include a 345~GHz receiver.  The largest simultaneous array that can
be deployed at 345~GHz in the near future consists of the three US
telescopes and a Chilean station.  Should phased ALMA not become
available, an array consisting of Hawaii, SMTO, CARMA, Chile~1, and
the LMT with a 345~GHz receiver (but not otherwise optimized for
345~GHz performance) would suffice to detect periodicity in Models A-E
at bit rates of 2~Gbit\,s$^{-1}$ or higher.

Observations have recently been taken with an array consisting of the
JCMT, SMTO, and CARMA-1 at 230~GHz \citep{doeleman08}.  We run
simulations of this array, which is usable already, as well as the
same three telescopes with the addition of Chile-1, since APEX is
likely to have 230~GHz capability in the near future.  We find that
periodicity may be just marginally detectable at the current maximum
bit rate of 4~Gbit\,s$^{-1}$ if the source geometry in Sgr~A* is
favorable.  Larger bandwidths are critical for detecting periodicity
on these arrays (and may not suffice if only the US telescopes are
used, depending on source geometry).  It is likely that phased-array
capability at Hawaii and CARMA will become available on the same time
scale as higher bit rate capability.  At 345~GHz, an array consisting
of the JCMT, SMTO, and Chile-1 (APEX or ASTE) would not suffice to
detect periodicity except at very high bit rates ($\geq
32$~Gbit\,s$^{-1}$, depending on the model).

\subsection{Long-Period Models}

\begin{figure}
\resizebox{\hsize}{!}{\includegraphics{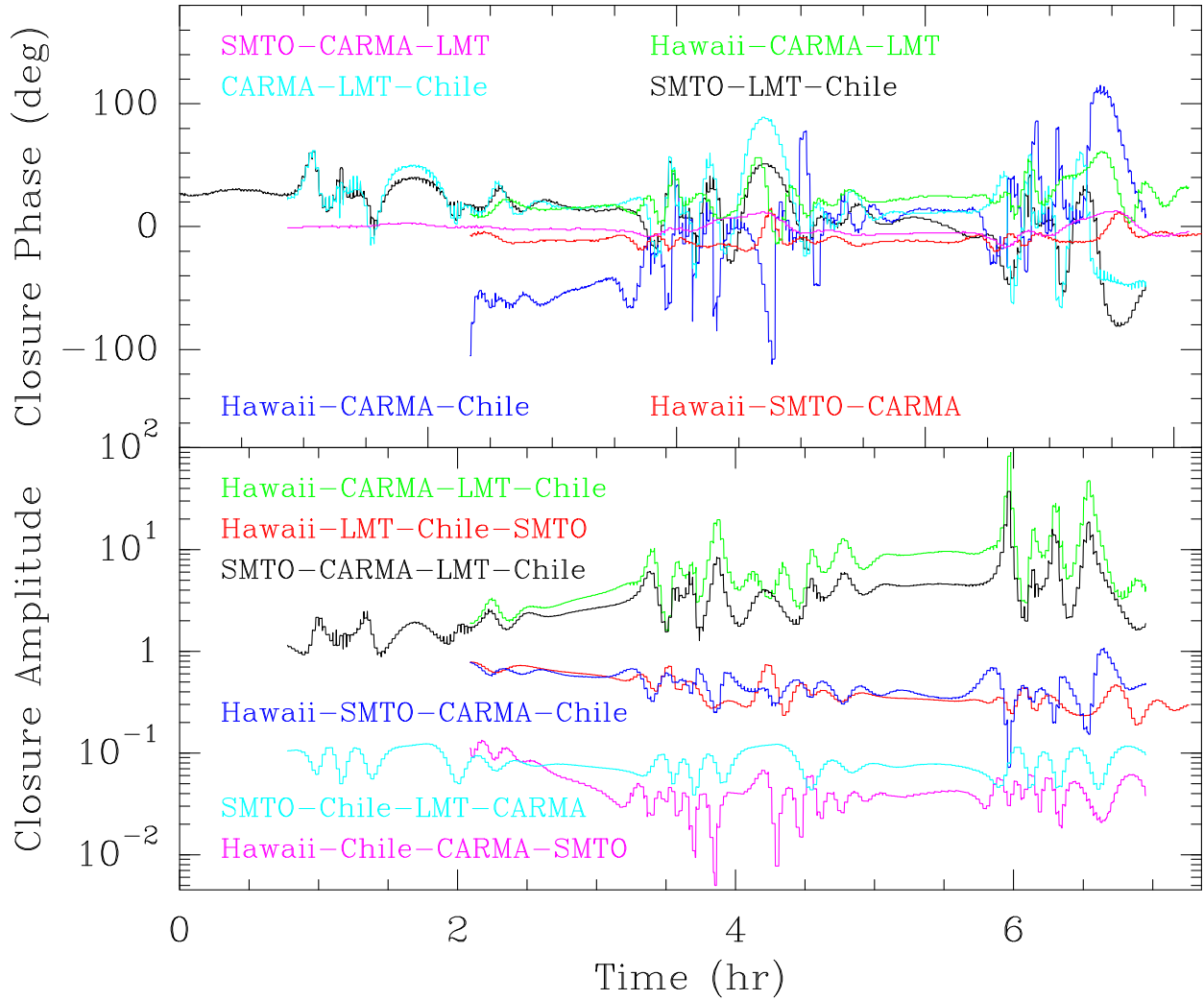}}
\caption{Closure phases and amplitudes for Model F at 230~GHz (period
  2.8~hr).  The colored lines indicate the closure quantities that
  would be obtained in the absence of noise.  The abscissa shows the
  time range over which Sgr~A* is above $5\degr$ elevation at at least
  three western hemisphere telescopes.  Due to changing baseline
  orientations, closure phases and amplitudes do not approximately
  repeat, although significant deviations are seen simultaneously on
  most subarrays.  Very small spikes seen in some of the lines are
  artifacts of the modelling.
  \label{fig-modelF}
}
\end{figure}

Detecting periodicity in the closure quantities will be more difficult
if the hot spot orbital period is long (e.g., several hours).  The
change in baseline orientation is significant enough within the
167-minute orbital period in Model F that the closure quantities do
not approximately repeat, as shown in Figure~\ref{fig-modelF}.
Consequently, the corresponding ACFs on most triangles and quadrangles
fall off with increasing lag, so the period cannot be determined by
finding the largest peak in the ACF.  Closure phases and amplitudes
show periods of relative quiescence punctuated by periods of large
simultaneous variability on most subarrays.

The mutual visibility on most triangles and quadrangles is comparable
to twice the orbital period of the long-period models, so periodic
behavior in closure phases and amplitudes may not be convincing in
demonstrating source structure periodicity.  For instance, it may not
be clear whether closure phases and amplitudes such as those shown in
Figure~\ref{fig-modelF} are due to a hot spot in a large orbit or due
to two unrelated flaring episodes of an aperiodic nature.  The LMT may
be critical for establishing long-period periodicity, since its
inclusion lengthens the total time range for observing, increasing the
chance that a third period will be detected.  While present
millimeter-VLBI arrays are thus not optimal for detecting long-period
orbiting hot spots, observational evidence suggests that hot spot
periods may be significantly shorter than that in Model~F.  VLBI
measurements of the position of the centroid of emission from Sgr~A*
presently constrain the orbital period of hot spots to be $\lesssim
120$~min if the hot spot flux dominates the disk flux, with
progressively longer periods allowed as the hot-spot-to-disk flux
ratio decreases \citep{reid08}.

However, there are fundamental constraints upon the existence of such
long-period hot spots.  The synchrotron cooling time at millimeter
wavelengths is roughly $3\,\hr$ (see eq.\,[\ref{eqn-cooling}]), and
thus in the absence of a continuous mechanism for injecting energetic
electrons in the hot spot, we would not expect to see single hot spots
persist longer than this.  Additionally, the available energy
(magnetic \& hydrodynamic) that may be tapped to generate substantial
emission decreases rapidly with radius, implying that the brightest
(and therefore dominant) events will preferentially occur at small
radii, with correspondingly short dynamical timescales.

\section{Discussion}
\label{discussion}

\subsection{Observing Strategies and Telescope Prioritization}

Ultimately, observations are limited by the telescopes that are
available.  Individual observers often have little influence over
telescope construction priorities.  Nevertheless, it is important to
consider the relative advantages of planned and potential instruments.
Future observers may be limited by scarce resources, such as DBEs and
recording equipment, instead of by available telescopes.

Array phasing will be crucial for maximizing the potential of
millimeter VLBI arrays.  Phased versions of CARMA and the millimeter
telescopes on Mauna Kea are necessary in order to increase the
detectability of very weak signals on the long baselines.  However,
further observations on the Hawaii-CARMA-SMTO triangle should not be
deferred due to the lack of a phased-array processor at Hawaii and/or
CARMA.  In some models, periodicity could be detected on the triangle
consisting of the JCMT, a single CARMA dish, and the SMTO at high
bandwidth.  Most of our models also indicate that periodicity may be
detectable on the Hawaii-CARMA-Chile~1 triangle, although the use of
phased ALMA rather than a single dish as the Chilean telescope may
also be important for detecting periodicity if the black hole is
highly rotating or if hot spot flux density signatures are typically
smaller than assumed in this work.

Given the schedule of proposed telescope upgrades, observations
utilizing closure techniques in the near future are most likely to use
the Hawaii-CARMA-SMTO triangle at 230~GHz and the Hawaii-SMTO-Chile
triangle at 345~GHz, with the possibility that the JCMT and/or CARMA-1
may necessarily be used in place of Hawaii and CARMA for the earliest
observations.  As illustrated in Figure~\ref{fig-phase230} (and
Table~\ref{tab-results}), the Hawaii-CARMA-SMTO triangle at 230~GHz
may not provide adequate spatial resolution to detect an orbiting hot
spot, depending on the parameters of the Sgr~A* system.  At the other
extreme, the currently available array of Hawaii, SMTO, and Chile-1 at
345~GHz may resolve out most of the emission (Fig.~\ref{fig-uv}) and
is unlikely to be useful for periodicity detection by itself at data
rates less than about 16~Gbit\,s$^{-1}$.  We therefore conclude that
observations of Sgr~A* in the near future to detect periodicity should
be taken at 230~GHz.

It will be important to include a fourth antenna in the array at each
frequency.  As demonstrated in \S \ref{ACF}, the additional closure
phases and amplitudes obtained by a 4-element array may be important
for detecting periodicity.  Even if periodicity can be marginally
detected by a 3-element array, the extra 4 independent closure
quantities provided by inclusion of a fourth telescope will vastly
increase the significance of periodicity detection.  Early
observations at 230~GHz should include APEX or a single ALMA dish, if
at all possible, in order to provide a fourth station in the array
(otherwise assumed to include Hawaii, SMTO, and CARMA).  A key benefit
of a 4-station US-Chile array, beyond the much larger number of
closure quantities provided, is that the constituent baselines will be
sensitive to a large range of spatial scales, making it less likely
that periodicity will be missed due to a mismatch between the angular
scale of source structure variability and the angular resolution of
the array.

The LMT will be a critical telescope for two reasons.  First, it fills
in an important hole in the $(u,v)$ plane, allowing for a large range
of angular scales to be probed by the full array.  Indeed, the LMT
produces excellent closure phase data on some triangles for every
model considered in this work.  Second, inclusion of the LMT allows
for simultaneous visibility of Sgr~A* from up to 5 telescopes.  The
utility of the LMT derives especially from its location and its size.
Even an incomplete LMT (e.g., a 32 m dish, as assumed in this work)
whose surface has not yet been fully tuned to maximize aperture
efficiency would be highly useful for observations of the Galactic
center at 230~GHz.  Thus, we conclude that millimeter-wavelength VLBI
arrays observing Sgr~A* should include the LMT as soon as is
practical.

There is also a need for 345~GHz capability at a greater number of
telescopes.  CARMA would be especially useful due both to its large
effective collecting area as well as its short baselines to the SMTO
and Hawaii.  The same comments that apply to the LMT in its planned
230~GHz band also apply to a future 345~GHz system, if ever planned.
Expanding PV to include 345~GHz would allow for observations on the
Europe-Chile triangle, which would provide only one additional closure
phase and is therefore not a high priority for periodicity detection.
However, the long Europe-Chile baselines cover an otherwise unsampled
region of the $(u,v)$ plane and may therefore prove to be important
for eventual modelling of the Sgr~A* quiescent disk.

An alternative observing strategy for the Chilean site, if phased ALMA
is unavailable, would be to use multiple individual telescopes.
Baselines from the Chilean telescopes (APEX, ASTE, or a single ALMA
dish) to North American or European telescopes would effectively be
redundant with each other in the $(u,v)$ plane but would offer
independent data, thus increasing the number of closure quantities
available.  Baselines between Chilean telescopes would be too short to
resolve the emission from Sgr~A* but would provide a valuable
simultaneous (near-)zero-spacing flux measurement, allowing estimation
of the fraction of flux resolved out by the short PV-PdB or SMTO-CARMA
baselines.

\subsection{Bandwidth Considerations}

Our simulations indicate that the periodicity of reasonable hot spot
models is detectable on most millimeter VLBI triangles/quadrangles at
data rates from 1 to 32~Gbit\,s$^{-1}$.  However, we note several
caveats whose applicability depends on the exact physical parameters
of the Sgr~A* system.  It may be difficult to detect periodicity on
the smallest triangles due to insufficient spatial resolution.
Conversely, the largest triangles may resolve out much of the
emission.  Very high data rates may be required to detect periodicity
on long triangles to Chile~1.  During any particular track of
observations, it is possible that no hot spot will be present, or that
the flux density of the hot spot may be substantially smaller than
assumed here.

We have assumed throughout that the atmospheric coherence time is 10~s.  In
poorer weather conditions, the coherence time may be significantly shorter.
Since the SNR of a coherently-integrated visibility grows as $(Bt)^{1/2}$,
where $B$ is the bandwidth, the SNR for a single visibility integrated over a
2.5~s coherent time interval will be half that of a visibility integrated over
10~s.  So long as the signal is strong enough compared to the noise (i.e., the
SNR $\gg 1$ in a single coherent integration), there is no net loss of signal
because the decrease in SNR is exactly compensated by the increase in the
number of data points obtained.  The SNR of the closure phase can then be built
up over a longer period of time, if desired, by averaging consecutive closure
phases (provided that the noiseless closure phase is not changing rapidly over
the time scale of integration).  Periodicity may still be weakly detected if
the average SNR exceeds unity over only a portion of the hot spot orbit, as
could be the case if the visibility amplitude on the weakest baseline changes
by a factor of several over an orbit.

Due to this effective SNR cutoff, it will be important to obtain data
at the maximum bandwidth (or recording rate) possible at the time of
observations.  Initial observations may be bandwidth-limited to 1~GHz
(4~Gbit\,s$^{-1}$).  If subsequent observations are limited by average
recording rate rather than IF bandwidth, it will be advantageous to
use burst-mode recording, if available.  In the limit of marginal
detections, it is far preferable to have a reduced number of good data
points rather than a full set of poor data.

\subsection{Black Hole Parameter Estimation}

The observation of periodicity in the closure quantities would provide
important evidence, independent of observations of periodicity in
flare light curves, that at least some subset of Sgr~A*'s flares
are due to bright orbiting structures.  This is critical to
efforts to use such structures ( e.g., hot spots) to infer the
properties of the black hole spacetime.

In principle, combined with the flare amplitude (as measured via the unresolved
light curves), the degree of variability in the closure quantities is
indicative of the size of the hot-spot orbit.  Combined with the period, this
provides a straightforward way in which to estimate the spin.  However, the
precision with which submillimeter closure quantities can constrain the spin
has yet to be determined, and will likely have to await a detailed
parameter-space study.  On the topic of black hole spin, it should be pointed
out that arguments for a non-zero spin of the SgrA* black hole can be made
based on the observed intrinsic size of Sgr~A* at 1.3~mm \citep{doeleman08},
and from NIR variability results \citep{genzel03}.

Our methods are generalizable to any mechanism of flare production.
There will almost surely be asymmetric structure on scales of a few to
a few tens of $\Rs$ due to general relativistic effects from the
accretion disk.  Regardless of whether flares are produced by orbiting
or spiralling hot spots, jets, magnetohydrodynamic instabilities, or
some other mechanism, there will also be asymmetric structures in the
inner disk region on scales of a few $\Rs$.  Our models
demonstrate that millimeter-wavelength VLBI arrays will be sensitive
to changes in the source structure no matter how these asymmetries are
oriented relative to each other on the sky.

\subsection{Summary of Simulation Findings}

\begin{itemize}

\item The currently useable 230~GHz array consisting of SMTO, CARMA-1, and JCMT
at 4~Gbit\,s$^{-1}$ cannot detect periodicity in any of the models tested.
\item Phasing connected element arrays (CARMA, Hawaiian telescopes, ALMA) to
form large effective apertures, significantly improves the probability of
detecting the hot spot period.  When the CARMA array is phased, and the
Hawaiian telescopes are coherently summed, periodicity in two out of five of
the 230~GHz models can be detected.  When 10 elements of ALMA are phased
together, periodicity can be reliably extracted in all of the 230~GHz models
tested.  
\item Higher recording bandwidth increases signal-to-noise on all baselines,
thereby improving periodicity detection, and making the array more robust
against loss of VLBI signal coherence due to atmospheric turbulence.  In three
out of five 230~GHz models tested, a recording bandwidth of 16~Gbit\,s$^{-1}$
allows detection of periodicity using just a three-station VLBI array.  
\item Adding a fourth or fifth telescope to the array enables detection of the
hot-spot period \emph{in every} 230~GHz model tested.  The added baseline
coverage in larger arrays allows more complete sampling of spatial scales in
SgrA*.  
\item Inclusion of the LMT in 230~GHz arrays will be very important as
baselines to the LMT fill critical voids in baseline coverage.  When long
baselines between the US and Chile heavily resolve SgrA*, baselines between
each of these regions and the LMT can still provide high signal-to-noise
detections that connect all telescopes in the array.  
\item A 345~GHz capability at more sites is needed.  The currently available
345~GHz array, consisting of SMTO, Hawaii, and Chile has long baselines that
will resolve much of the flux density of SgrA*.  Enhancing either CARMA or the
LMT with low-noise 345~GHz receivers, puts periodicity detection of all the
tested 345~GHz models within reach when ALMA comes on line.

\end{itemize}

\subsection{Conclusions}

Motivated by recent detection of intrinsic structure within SgrA* on $<4\Rs$
scales, this work explores the feasibility of detecting time variable structure
and periodicity in the context of flare models in which a hot-spot orbits the
central black hole.  Algorithms are described that use interferometric closure
quantities, direct VLBI observables which reflect asymmetries in source
structure that can be tracked on time scales that are short compared to
presumed hot-spot orbital periods.  We find that periodicity from these models
over a representative range of parameters can be reliably extracted using mm
and submm VLBI arrays that are planned over the next 5 years.  Thus, mm and sub-mm
VLBI has matured to the level where one can envisage studying fundamental black
hole parameters, accretion physics, and General Relativity in the strong field
regime at meaningful angular resolutions.

The techniques and concepts described here are applicable to a broad range of
Sgr~A* models, including emission due to jets and outflows, MHD simulations,
and adiabadically expanding flaring structures.  Closure amplitude analysis,
for example, is capable of tracking the size of SgrA* over time and sensitively
testing for expansion during a flare event.  Short wavelength VLBI, when
coupled with wide spectrum simultaneous monitoring, can thus make detailed
tests of emission models for Sgr~A* in which low frequency synchrotron photons
are up-scattered to produce X-rays.

The three fundamental technical efforts to improve mm and submm arrays include
increasing the number of VLBI sites, achieving the capability to phase
connected element arrays into a large effective aperture, and increasing the
VLBI recording data rate.  Projects in each of these areas, supported by an
international collaboration, are underway, and it is expected that observations
on VLBI arrays at 1.3~mm and 0.85~mm wavelength with dramatically improved
sensitivities will begin by 2009.  The increase of bandwidth, while of
general importance to lowering VLBI detection thresholds, may be most useful by
enabling full polarization observations.  The techniques developed in this work
can be extended to non-imaging VLBI polarimetry analysis, allowing tests for
small scale polarization structure to be carried out.  The prospect of using
VLBI to probe such polarized sub-structure emphasizes the power of the
technique for studying Sgr~A*.

\acknowledgments

The high-frequency VLBI program at Haystack Observatory is funded
through a grant from the National Science Foundation.

\end{document}